\documentclass[review,onefignum,onetabnum]{article}

\usepackage{latexsym}
\usepackage{amssymb,amsbsy,amsmath,amsfonts,amssymb,amscd,amsthm}
\usepackage{graphicx}
\usepackage{hyperref}
\usepackage{dsfont}
\usepackage{mathtools}
\usepackage{pdfsync}
\usepackage{epsfig,epstopdf}
\usepackage{xcolor}
\usepackage{hyperref}
\usepackage[title]{appendix}
\usepackage{verbatim}
\usepackage{lipsum}
\usepackage{float}
\usepackage[font=small]{caption}
\usepackage{subcaption}

\newtheorem{rmrk}[]{Remark}

\DeclarePairedDelimiter {\abs}{\lvert}{\rvert}

\newcommand{\R}{\mathbb{R}}
\newcommand{\N}{\mathbb{N}}
\newcommand{\Z}{\mathbb{Z}}
\newcommand{\di}{\,\mathrm{d}}
\newcommand{\pt}{\partial_t}
\newcommand{\px}{\partial_x}
\newcommand{\pxx}{\partial_{xx}^2}
\newcommand{\pr}{\partial_r}

\newcommand{\beq}{\begin{equation}}
	\newcommand{\eeq}{\end{equation}}
\newcommand{\beqa}{\begin{eqnarray}}
	\newcommand{\eeqa}{\end{eqnarray}}

\newcommand\bigO{
	\mathchoice
	{{\mathcal{O}}}
	{{\mathcal{O}}}
	{{\scriptstyle\mathcal{O}}}
	{\scalebox{.7}{$\scriptstyle\mathcal{O}$}}
}

\providecommand{\keywords}[1]
{
	\small	
	\textbf{\textit{Keywords---}} #1
}

\renewcommand{\vec}{\boldsymbol}

\numberwithin{equation}{section}
\numberwithin{prpstn}{section}
\numberwithin{ass}{section}
\numberwithin{rmrk}{section}

\title{Agent-based and continuum models for spatial dynamics of infection by oncolytic viruses\thanks{Corresponding author: David Morselli (david.morselli@polito.it; dmorselli@swin.edu.au)\\
		This research was partially supported by the Italian Ministry of Education, University and Research (MIUR) through the “Dipartimenti di Eccellenza” Programme (2018-2022) – Dipartimento di Scienze Matematiche “G. L. Lagrange”, Politecnico di Torino (CUP: E11G18000350001). MED and DM are members of GNFM (Gruppo Nazionale per la Fisica Matematica) of INdAM (Istituto Nazionale di Alta Matematica). We also acknowledge the support of the Australian National Health and Medical Research Council, through grant NHMRC IDEAS 2013058. Part of this work was performed on the OzSTAR national facility at Swinburne University of Technology. The OzSTAR program receives funding in part from the Astronomy National Collaborative Research Infrastructure Strategy (NCRIS) allocation provided by the Australian Government, and from the Victorian Higher Education State Investment Fund (VHESIF) provided by the Victorian Government.}
}
\author{David Morselli,\thanks{Department of Mathematical Sciences ``G. L. Lagrange'', Politecnico di Torino, Corso Duca degli Abruzzi 24, 10129 Torino, Italy} \thanks{Department of Mathematics, School of Science, Computing and Engineering Technologies, Swinburne University of Technology, John St, 3122, Hawthorn, VIC, Australia} \thanks{Department of Mathematics ``G. Peano'', Università di Torino, Via Carlo Alberto 10, 10124 Torino, Italy}
	\and
	Marcello Edoardo Delitala,\footnotemark[2]
	\and
	Federico Frascoli\footnotemark[3] 
}

\date{}

\begin{document}
	
	\maketitle
	
	\begin{abstract}
		The use of oncolytic viruses as cancer treatment has received considerable attention in recent years, however the spatial dynamics of this viral infection is still poorly understood. We present here a stochastic agent-based model describing infected and uninfected cells for solid tumours, which interact with viruses in the absence of an immune response. Two kinds of movement, namely undirected random and pressure-driven movements, are considered: the continuum limit of the models is derived and a systematic comparison between the systems of partial differential equations and the individual-based model, in one and two dimensions, is carried out.
		
		In the case of undirected movement, a good agreement between agent-based simulations and the numerical and well-known analytical results for the continuum model is possible. For pressure-driven motion, instead, we observe a wide parameter range in which the infection of the agents remains confined to the center of the tumour, even though the continuum model shows traveling waves of infection; outcomes appear to be more sensitive to stochasticity and uninfected regions appear harder to invade, giving rise to irregular, unpredictable growth patterns. 
		
		Our results show that the presence of spatial constraints in tumours' microenvironments limiting free expansion has a very significant impact on virotherapy. Outcomes for these tumours suggest a notable increase in variability. All these aspects can have important effects when designing individually tailored therapies where virotherapy is included.
		
	\end{abstract}
	
	
\section{Introduction}

Oncolytic viruses constitute a targeted cancer therapy, that uses viral particles preferentially infecting tumour cells while mostly sparing healthy tissues \cite{blanchette23,fountzilas17,Kelly2007651review,lawler17,russell18}. Although the potential of this therapy has been stressed for a long time, the clinical use still faces many challenges; one of them is the lack of understanding of the tumour microenvironment's role in viral diffusion \cite{jin21,wojton10}. The difficulties in creating a set of rules and practises that make this therapy reliable, reproducible and clinically mainstream are generally associated with ``stochastic'', hard-to-predict events, that affect consistency in viral delivery, tumour invasion, viral replication and diffusion. 

Several mathematical models have previously been adopted for the study of oncolytic viruses, including ordinary differential equations (ODEs) \cite{jenner19,jenner18insights,komarova10,novozhilov06,wodarz01}, partial differential equations (PDEs) \cite{alzahrani19,friedman03,kim14,pooladvand21,wu01,wu04}, stochastic agent-based models \cite{jenner20voronoi,wodarz12} and hybrid discrete-continuous multi-scale models \cite{jenner22,paiva09}. In \cite{wodarz16} a review of the different modeling approaches is presented. As it is well-known, individual-based models allow to track individual cells and consider randomness in the processes, but are also associated to higher computational cost and do not allow to easily obtain analytical results. On the other hand, deterministic continuum models are amenable both to numerical simulations and analytical results, but cannot easily include stochastic events; furthermore, the phenomenological assumptions commonly used in this approach may also hinder the biological interpretation of the mathematical assumptions. For these reasons, in recent years the derivation of continuum macroscopic models from underlying discrete stochastic models has attracted the attention of an increasing number of researchers (see, for example,  \cite{champagnat07,johnston15,lorenzimurray20, macfarlane22,penington11}; we refer to the introduction of \cite{chaplain20} for a more comprehensive literature review). This allows to understand clearly the modeling assumptions for a continuum model, gain some theoretical intuition on the behavior of an individual-based model and, as a consequence, reach a more comprehensive understanding of the biological system under study. 

With the exception of \cite{wodarz12}, we are not aware of any other work comparing agent-based and continuous models in relation to oncolytic viruses. In this paper we bridge such a gap by developing an original, minimal spatial individual-based model for the infection of tumour cells due to engineered viruses. Our model takes into account proliferation and death of uninfected tumour cells, death of infected tumour cells, infection of uninfected cells and cell movement. We present two alternative sets of rules governing the latter process (namely, undirected random cell movement and pressure-driven cell movement \cite{chaplain20}) and show how this choice strongly influences therapy outcomes. Our intent is to compare different mechanisms for tumour development, capturing some of the constraints that diverse microenvironments pose on tumours' development. Viral responses and pattern of invasion appear to be clearly affected, often in unpredictable ways.

While it is known that oncolytic viruses are able to infect through specific receptors that are highly expressed on cancer cells \cite{lawler17}, the exact mechanisms of the infection are not well understood. Viruses enter target cells with a combination of dynamics, whose effectiveness depends on a number of factors \cite{kalia11}. In recent years it has also become clear that some viruses (such as human immunideficency virus type 1 and hepatitis C virus) may infect both through direct cell-to-cell trasmission and cell-free trasmission mediated by diffusing virions; the actual combination of the two processes is hard to establish in full detail (see \cite{graw16} and the references therein). There are also newly investigated mechanisms that allow cell-to-cell transmission: for example, some viruses such as influenza virus exploit tunneling nanotubes between cells \cite{kumar17}. Since all these dynamics for oncolytic viruses are mostly unknown, for the sake of simplicity we assume that the infection happens when an uninfected cell has a contact with an infected cell and viral spread far from infected cells can be neglected \cite{wodarz12}. This approach has been commonly used for nonspatial models of oncolytic viruses \cite{komarova10,novozhilov06}. In the context of spatial models, this choice allows to model a virus that faces some difficulties in propagating in the tumour microenvironment \cite{wojton10} and thus the infection is mainly driven by cell-to-cell contact and close range free virions. Similarly, we do not include virus clearance due to the immune system and assume that no immune response is present. The dynamics of viruses and immune system are somewhat implicitly taken into account in the definition of the infection rate (see Appendix \ref{app:num}), although clearly the influence of the immune system is much more complicated and its analysis goes beyond the scope of the present work. Finally, we postulate that a limited viral load is injected at the centre of the tumour, in line with typical clinical practices \cite{russell18}.

The resulting systems fall in the category of classical spatial Lotka--Volterra models for preys and predators. In the ecological setting the comparison between discrete and continuum models of this form has been widely studied (for example, in \cite{aronson80,keeling02,wilson93}). In the case of undirected cell movement, the corresponding continuum model is a diffusive Lotka--Volterra model with logistic growth: this allows us to partially rely on previous analytical results on the subject \cite{dunbar84}. On the other hand, in the case of pressure-driven cell movement the corresponding continuum model is a local cross-diffusion Lotka--Volterra model that we could not find in the literature (although it is similar to the systems studied in \cite{bubbaheleshaw20,carrillo18,gwiazda19,lorenzi17}). Our results suggest that stochastic events may hinder the propagation of the infection even in situations in which the continuous model shows the formation of a traveling infection wave.

The article is organised as follows. In Section \ref{sec:model}, we introduce the two agent-based models and present their continuum counterpart (a formal derivation is presented in Appendix \ref{app:derivation}). In Section \ref{sec:wave} we present some classical analytical results for traveling wave solutions of the continuum models. In Sections \ref{sec:undirected} and \ref{sec:pressure} we compare the results of numerical simulations of the two agent-based models and the numerical solutions of the corresponding PDEs, showing their consistency with the analytical results. In Section \ref{sec:conclusion}, we discuss the main findings in light of existing experimental evidence \textit{in vitro} and provide some hints for future research.

\section{Description of the agent-based models and formal derivation of the continuum models}
\label{sec:model}

In this section we describe the stochastic dynamics of the two agent-based models and introduce the different expressions of the probability for the cell movement. We then present the corresponding continuum counterparts, obtained in Appendix \ref{app:derivation} using techniques analogous to those employed in various references~\cite{champagnat07,johnston15, lorenzimurray20,macfarlane22,penington11,chaplain20,almeida22,almeida23}.

\subsection{Agent-based models}
In the agent-based modeling framework, each cell is an agent occupying a position on a discrete lattice. We consider two cell populations, uninfected and infected; the infection of a cell then corresponds to an agent passing from the former to the latter population. Cells can also move, reproduce and die. For ease of presentation, in this section we only consider cells arranged along the one-dimensional real line $\R$, but there would be no additional difficulty in considering higher spatial dimensions. Since we carry out the comparisons between discrete and continuum models also in two spatial dimensions, in Remarks \ref{rmrk:dif_twod} and \ref{rmrk:pres_twod} we explain the small changes of the two-dimensional models.

Let us consider the temporal discretisation $t_n=\tau n$ with $n\in\N_0$, $0<\tau \ll 1$ and the spatial discretisation $x_j=\delta j$, with $j\in\Z$, $0<\delta\ll 1$; we assume $\tau$ to be small enough to guarantee that all the probabilities defined hereafter are smaller than $1$. We denote the number of uninfected and infected cells that occupy position $x_j$ at time $t_n$ respectively by $U_j^n$ and $I_j^n$; the corresponding densities are
\[
u_j^n\coloneqq \frac{U_j^n}{\delta}, \qquad i_j^n\coloneqq \frac{I_j^n}{\delta}
\]
The local pressure is assumed to be given by a barotropic relation of the form $\rho_j^n\coloneqq \Pi(u_j^n+i_j^n)$. In the next sections we restrict to the case $\Pi(z)=z$ (so the pressure is actually the total cell density), but the discussion of this section is valid also for more general nondecreasing smooth functions $\Pi\colon[0,+\infty)\to[0,+\infty)$ such that $\Pi(0)=0$: for example, one could think of the functional form proposed in \cite{perthame14}. Although there might be differences in the way the system reaches the carrying capacity and how the model appears in the continuum limit, it seems that the overall behaviour of the tumour is not profoundly affected by different functional forms \cite{macfarlane22}.

\begin{figure}[h]%
	\centering
	\includegraphics[width=\textwidth]{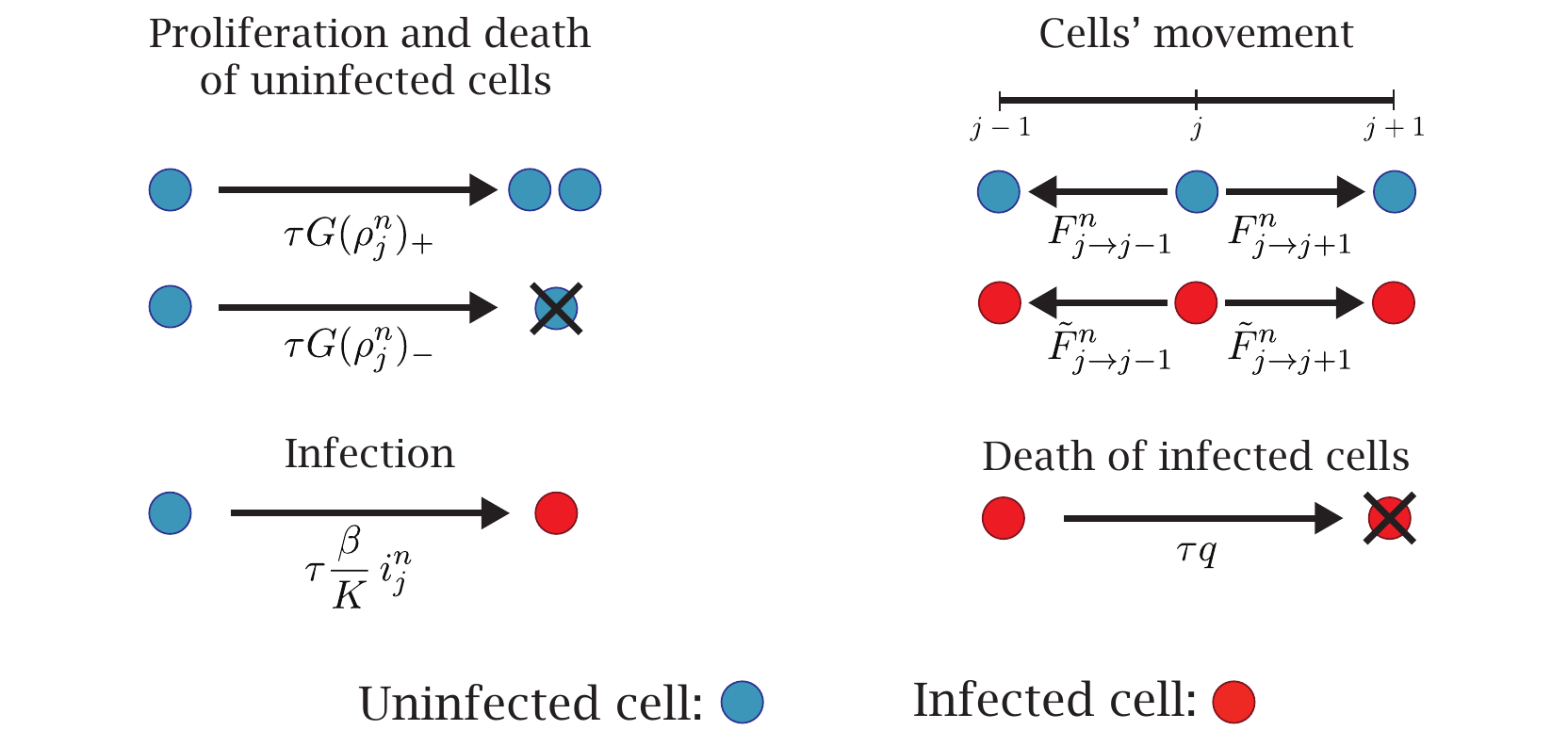}
	\caption{Schematic representation of the rules governing cell dynamics in the stochastic models. Uninfected cells are represented in blue and infected cells in red. Uninfected cells may proliferate or die according to the pressure value, move and become infected upon contact with infected cells. Infected cells may move and die with constant probability. We consider different expressions for the probabilities of movement, given respectively in Eqs. \eqref{eq:movement_undirected} and \eqref{eq:movement_pressure}.}
	\label{fig:discrete}
\end{figure}

Fig. \ref{fig:discrete} summarises the rules governing the dynamics of the agents. We consider two different movement mechanism, i.e. undirected and pressure-driven, giving rise to different models. The rules for proliferation and death of uninfected cells, death of infected cells and infection are common for both models.

\paragraph{Pressure-dependent proliferation of uninfected cells} We assume that the proliferation probability decreases as the pressure increases and stops at some homeostatic pressure $P>0$; a pressure value greater than $P$ results in the cell's death. Given a smooth decreasing function $G\colon [0,+\infty)\to\R$ such that $G(P)=0$, we let an uninfected cell that occupies position $x_j$ at time $t_n$ reproduce with probability $\tau G(\rho_j^n)_+$, die with probability $\tau G(\rho_j^n)_-$, and remain quiescent with probability $1-\tau G(\rho_j^n)_+-\tau G(\rho_j^n)_-=1-\tau \abs{G(\rho_j^n)}$. In these formulas, $z_+\coloneqq\max\{z,0\}$ and $z_-\coloneqq\max\{-z,0\}$. When a reproduction takes place, a new cell is placed at the same lattice site. The fact that proliferation stops above $P$ guarantees that, as $\tau\to 0$, a population of cells whose initial pressure is below the homeostatic value becomes less likely to acquire a pressure value above this level at later times. This kind of probabilities has already been employed in \cite{chaplain20}.

For the sake of simplicity, in the following sections we restrict our analysis to the logistic growth, i.e.
\begin{equation}
	\label{eq:G}
	G(\rho)=p\Bigl(1-\frac{\rho}{P}\,\Bigr)
\end{equation}
where $p>0$ is the maximal duplication rate. Let us observe that the carrying capacity of the system is $K\coloneqq \Pi^{-1}(P)$; since in the case of our interest $\Pi(z)=z$, we actually have $P=K$.

\paragraph{Death of infected cells} We do not model proliferation of infected cells, as the virus disrupts the cellular machinery. Some time after the infection, the cell undergoes lysis and dies: we assume that at every time step this happens with probability $\tau q$, where $q>0$ is a constant death rate.

\paragraph{Infection} We do not model explicitly the oncolytic virus, as we assume that its dynamics are faster than cellular dynamics and can thus be approximated by a quasi-steady state (as in  \cite{komarova10,novozhilov06}; see also Appendix \ref{app:num}). Thus, we assume that infection takes place upon contact between infected and uninfected cells with probability proportional to the density of infected cells. This means that an uninfected cell that occupies position $x_j$ at time $t_n$ becomes infected with probability $\tau\beta i_j^n/K$, where $K$ is the carrying capacity and $\beta>0$ is a constant death rate. Although the carrying capacity could be easily incorporated in the infection parameter, this formulation allows to easily rescale the cell densities by only modifying $K$ and the initial conditions. This process is similar to the interaction between the tumour and the immune system described, for example, in \cite{almeida22,almeida23}.

\paragraph{Cell movement} As we already mentioned, we consider two different rules governing cell movement. In view of the formal derivation of the continuum models, it is convenient to adopt the same notation for both processes. We thus state that an uninfected cell that occupies position $x_j$ at time $t_n$ moves to the lattice point $x_{j\pm 1}$ with probability $F_{j\to j\pm 1}^n$ and remains at its initial position with probability $1-F_{j\to j- 1}^n-F_{j\to j+ 1}^n$. The same happens for the infected cells, but with probabilities $\tilde{F}_{j\to j\pm 1}^n$ that in principle may be different from $F_{j\to j\pm 1}^n$.

Let us now give the explicit expressions for these probabilities. The simplest model of movement assumes no influence of the cell density and no preferential direction of motion; in this case we set
\begin{equation}
	\label{eq:movement_undirected}
	F_{j\to j\pm 1}^n\coloneqq\frac{\theta_u}{2}, \qquad \tilde{F}_{j\to j\pm 1}^n\coloneqq\frac{\theta_i}{2}
\end{equation}
with $\theta_u, \theta_i\in[0,1]$. This is a standard unbiased random walk.

On the other hand, since cellular proliferation is limited by a carrying capacity, it also makes sense to take into account a reduction of motility in crowded environment and allow cells to only move following the pressure gradient: the probability of movement thus depends on the difference between the pressure at the initial position of the cell and the pressure at the target point. In this case we set
\begin{equation}
	\label{eq:movement_pressure}
	F_{j\to j\pm 1}^n\coloneqq\theta_u \frac{(\rho_j^n-\rho_{j\pm 1}^n)_+}{2P}, \qquad \tilde{F}_{j\to j\pm 1}^n\coloneqq\theta_i \frac{(\rho_j^n-\rho_{j\pm 1}^n)_+}{2P}
\end{equation}
where $z_+\coloneqq\max\{z,0\}$, $P$ is the homeostatic pressure and $\theta_u, \theta_i\in[0,1]$ as before. Observe that, if $\rho_j^n\leq P$ for every $j$, then all the probabilities are between $0$ and $1$. This kind of reasoning and the probabilities associated have already been employed in \cite{chaplain20}. 

In the special case $\rho_j^n=P$ and $\rho_{j-1}^n=\rho_{j+1}^n=0$ the two definitions give the same probability values; in any other case, the probabilities of movement given in Eq. \eqref{eq:movement_undirected} are higher than the ones given in Eq. \eqref{eq:movement_pressure}. This, as we will see shortly, strongly affects the therapy outcomes.

\subsection{Continuum model in the case of undirected movement}
Here we consider the undirected cell movement with the probabilities given in Eq. \eqref{eq:movement_undirected}.
Letting $\tau,\delta\to 0$ in such a way that $\frac{\delta^2}{2\tau}\to D$ and assuming that there are two functions $u\in C^2([0,+\infty),\R)$ such that $u_{j}^{n}=u(t_n,x_j)$ and $i\in C^2([0,+\infty),\R)$ such that $i_{j}^{n}=i(t_n,x_j)$, we formally obtain (see Appendix \ref{app:derivation}) the following system of reaction-diffusion PDEs
\begin{equation}
	\begin{cases}
		\pt u(t,x)=D_u\pxx u(t,x)+pu(t,x)G(\rho(t,x))-\dfrac{\beta}{K} u(t,x)i(t,x)\\[8pt]
		\pt i(t,x)=D_i\pxx i(t,x)+\dfrac{\beta}{K} u(t,x)i(t,x)-qi(t,x)
	\end{cases}
\end{equation}
where $D_u\coloneqq \theta_u D$ and $D_i\coloneqq \theta_i D$.

If we take the function $G$ as in Eq. \eqref{eq:G} and $\rho=u+i$, then the system becomes
\begin{equation}
	\label{eq:l}
	\begin{cases}
		\pt u=D_u\pxx u+pu\Bigl(1-\dfrac{u+i}{K}\Bigr)-\dfrac{\beta}{K} ui \\[8pt]
		\pt i=D_i\pxx i+\dfrac{\beta}{K} ui-qi
	\end{cases}
\end{equation}
This model is a simplified version of the one studied in \cite{pooladvand21}, as here we do not consider viral dynamics explicitly. A similar diffusive Lotka--Volterra model with logistic growth has been studied in \cite{dunbar84}; it is important to observe that in our case the infected cells, which play the role of predators, contribute to the saturation of the growth of uninfected cells, which play the role of preys, hence Eq. \eqref{eq:l} cannot be adimensionalised exactly in the same way as the model in \cite{dunbar84}.
	
\begin{rmrk}
	\label{rmrk:dif_twod}
	When the spatial domain is the two-dimensional real plane $\R^2$ instead of the one-dimensional real line $\R$, the scalar index $j\in\Z$ should be replaced by the vector $\vec{j}=(j_x,j_y)\in\Z^2$ and the probability that a cell moves to one of the four neighboring lattice points is $\theta_k/4$, with $k=u,i$. We then need to scale $\tau$ and $\delta$ in such a way that $\frac{\delta^2}{4\tau}\to D$.
\end{rmrk}

\subsection{Continuum model in the case of pressure-driven movement}
Let us consider the pressure-driven cell movement with the probabilities given in Eq. \eqref{eq:movement_pressure}. Letting $\tau,\delta\to 0$ in such a way that $\frac{\delta^2}{2\tau}\to D$ and assuming that there are two functions $u,i$ as in the previous model, we formally obtain (see Appendix \ref{app:derivation}) the following local cross-diffusion system
\begin{equation}
	\begin{cases}
		\pt u(t,x)=\dfrac{D_u}{P} \px [u(t,x)\px \rho(t,x)]+pu(t,x)G(\rho(t,x))-\dfrac{\beta}{K} u(t,x)i(t,x) \\[8pt]
		\pt i(t,x)=\dfrac{D_i}{P} \px [i(t,x)\px \rho(t,x)]+\dfrac{\beta}{K} u(t,x)i(t,x)-qi(t,x)
	\end{cases}
\end{equation}
where $D_u\coloneqq \theta_u D$ and $D_i\coloneqq \theta_i D$. This model can be thought as the natural generalisation to infections of the model presented in \cite{perthame14,byrne09}. A similar system is studied in \cite{gwiazda19}, although it is important to remark that our infection term does not fit in the framework of reaction terms considered in that paper.

If we take the function $G$ as in Eq. \eqref{eq:G} and $\rho=u+i$ (so that also $P=K$), then the system becomes
\begin{equation}
	\label{eq:p}
	\begin{cases}
		\pt u=\dfrac{D_u}{K} \px [u\px (u+i)]+pu\Bigl(1-\dfrac{u+i}{K}\Bigr)-\dfrac{\beta}{K} ui \\[8pt]
		\pt i=\dfrac{D_i}{K} \px [i\px (u+i)]+\dfrac{\beta}{K} ui-qi
	\end{cases}
\end{equation}

\begin{rmrk}
	\label{rmrk:pres_twod}
	When the spatial domain is the two-dimensional real plane $\R^2$ instead of the one-dimensional real line $\R$, the scalar index $j\in\Z$ should be replaced by the vector $\vec{j}=(j_x,j_y)\in\Z^2$ and the probability that a cell moves to one of the four neighbouring lattice points is 
	\[
	\theta_k \frac{(\rho_{\vec{j}}^n-\rho_{\vec{j}+\vec{e}}^n)_+}{4P}
	\]
	with $k=u,i$ and $\vec{e}\in\{(\pm 1,0),(0,\pm 1)\}$. As in the case of Remark \ref{rmrk:dif_twod}, we need to scale $\tau$ and $\delta$ in such a way that $\frac{\delta^2}{4\tau}\to D$.
\end{rmrk}

\section{Traveling waves for the continuum models}
\label{sec:wave}
In view of the forthcoming comparison of the different models, it is useful to keep in mind some well-known analytical results about traveling waves. We can also anticipate that analytical results are still not available for the pressure-driven regime, although some numerical simulations, as we will see, work well. We first recall that the equation
\begin{equation}
	\label{eq:fisher}
	\pt u= D\pxx u+ pu\Bigl(1-\frac{u}{K}\Bigr)
\end{equation}
admits as solutions traveling waves with speed at least $2\sqrt{Dp}$ and an initial condition with compact support evolves into a wave that travels with the minimal speed \cite{fisher37,kolmogorov37}.
The application of standard linearisation techniques \cite[\S 2.1]{van03} yields the same invasion speed $2\sqrt{D_u p}$ for any reaction-diffusion equation $\pt u= D\pxx u+ f(u)u$ such that $f'(0)=p$. A special case is $f(u)$ constant and equal to $p$, which corresponds to exponential growth.

On the other hand, the equation
\begin{equation}
	\label{eq:nonlin_diff}
	\pt u= \frac{D}{K} \px (u\px u)+ pu\Bigl(1-\frac{u}{K}\Bigr)
\end{equation}
admits as solutions traveling waves with speed at least $\sqrt{Dp/2}$ and again an initial condition with compact support evolves into a wave that travels with the minimal speed \cite{aronson80,newman80}; the main difference with respect to the previous equation is the fact that initial data with compact support evolve into sharp waves with compact support.  It is important to observe that, since the spatial dependence is intrinsically nonlinear, a direct application of linear spreading speed does not give any meaningful information.

Let us also recall that the system
\begin{equation}
	\label{eq:ode_l}
	\begin{cases}
		\dfrac{\di u}{\di t}=pu\Bigl(1-\dfrac{u+i}{K}\Bigr)-\dfrac{\beta}{K} ui \\[8pt]
		\dfrac{\di i}{\di t}=\dfrac{\beta}{K} ui-qi
	\end{cases}
\end{equation}
(which is the spatially homogeneous analog of Eqs. \eqref{eq:l} and \eqref{eq:p}) has three equilibria: $(0,0)$, $(K,0)$ and 
\begin{equation}
	\label{eq:eq_l}
	(u^*,i^*)\coloneqq \Bigl( \frac{qK}{\beta}, \frac{pK(\beta -q)}{\beta(\beta +p)}\Bigr)
\end{equation}
The first equilibrium has eigenvalues $p$ and $-q$, so it is unstable (recall that all the parameters are strictly positive). The second one has eigenvalues $-p$ and $\beta -q$, so it is stable when $\beta<q$ (i.e., $i^*<0$) and unstable when $\beta>q$ (i.e., $i^*>0$). The expression for the eigenvalues of the last equilibrium is more complicated, but their sum is $-\frac{pq}{\beta}$ and their product is $\frac{pq(\beta-q)}{\beta }$: hence, when $i^*>0$ the eigenvalues are either both real and negative or complex with negative real part; in both cases, the equilibrium is stable. Observe that, in the case of $\beta<q$ and positive initial data, the only possible outcome is the extinction of infected cells and the growth of the uninfected cells to the carrying capacity, which in our biological interpretation corresponds to a complete failure of the treatment. As also pointed out in other works by some of the present authors, the interplay between infection rate and death rate of infected cells is responsible, to some extent, to the success of the overall therapy \cite{jenner18insights,pooladvand21}. Infections that start and develop too quickly seem to carry less ability to effectively control the tumour in the long run. This can be circumvented, to some extent, by encasing the virus in gels or implementing strategies to retard and prolong its release \cite{jenner19,pooladvand21,jenner20voronoi,jenner18biomath}. From now on we focus on the situation $\beta>q$.

\begin{figure}
	\centering
	\includegraphics[width=0.7\linewidth]{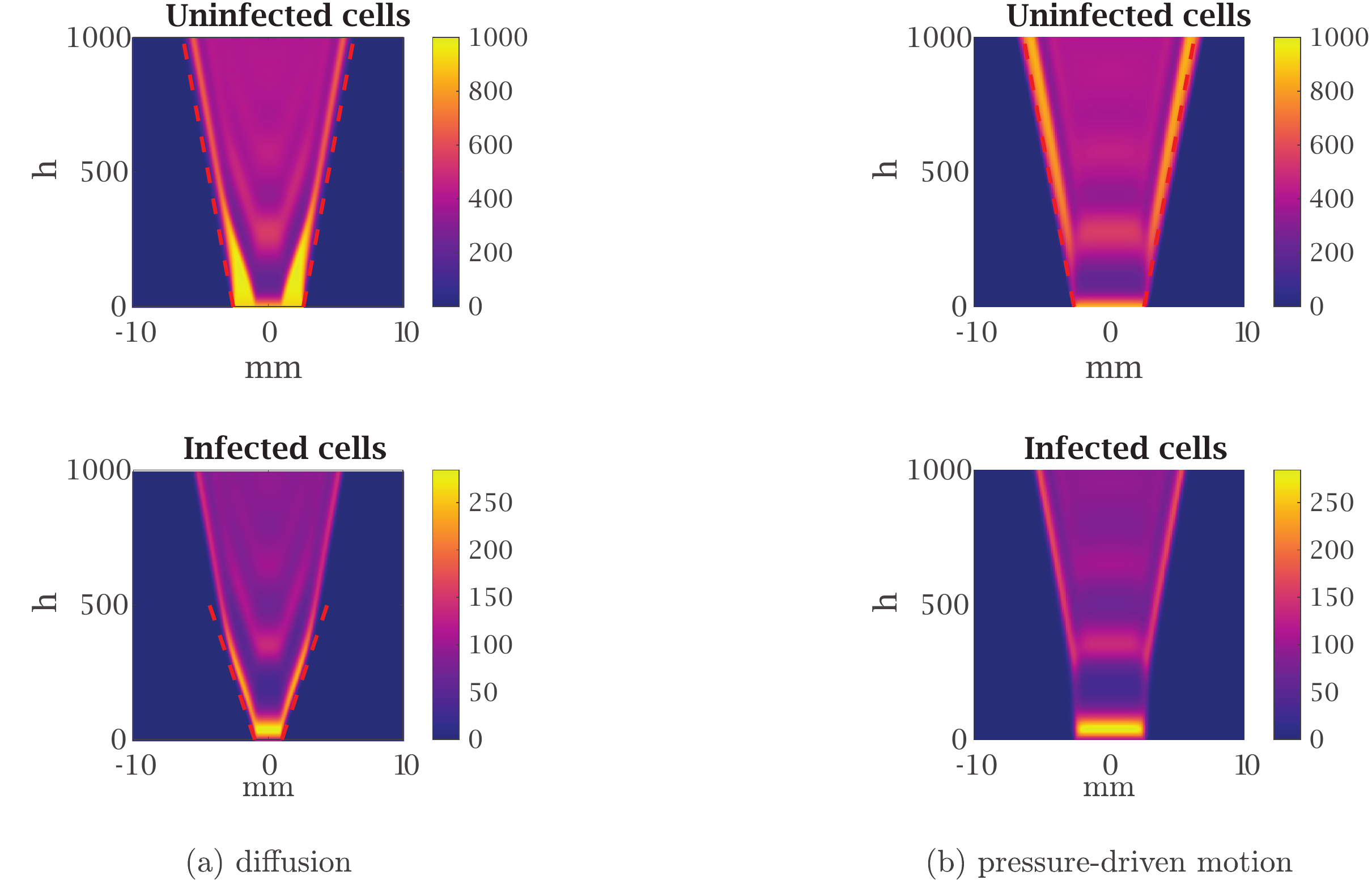}
	\caption{Numerical solutions of (a) Eq. \eqref{eq:l} and (b) Eq. \eqref{eq:p} show that theoretical results correctly estimate the speeds of the traveling waves. The results are discussed more in depth in the following sections. The parameters employed are the ones given in Table \ref{tab:parameters}, with the exception of the infection radius $R_i$ in panel (b) (which is set to $2.6\;$mm in order to allow the infection to spread).}
	\label{fig:speed}
\end{figure}

It is worth recalling that in \cite{dunbar84} it was proven that a system similar to Eq. \eqref{eq:l} (in which ``predators'' do not contribute to the saturation of uninfected cells' growth) admits traveling waves connecting $(K,0)$ to $(u^*,i^*)$ with speed at least $2\sqrt{D_i(\beta-q)}$ and damped oscillations after the front of the wave may appear. Since we are mostly interested in the case of a tumour that is still expanding, it makes more sense to look for traveling waves connecting $(0,0)$ to $(u^*,i^*)$ and we expect to observe a race between the uninfected cells (evolving according to Eq. \eqref{eq:fisher} in absence of infected cells) and the infected cells at the center of the tumour. This situation is shown in Fig. \ref{fig:speed}a and it is clearly more complex than the one examined in \cite{dunbar84}. Let us observe that the density of a population of infected cells invading a region of uninfected cells at constant density $\hat{u}$ satisfies the equation
\[
\pt i= D_i\pxx i+ \Bigl(\frac{\beta}{K}\hat{u}-q\Bigr)i
\]
This equation is analogous to the linearised version of Eq. \eqref{eq:fisher}, therefore we expect infected cells to travel at speed $2\sqrt{D_i(\frac{\beta}{K}\hat{u}-q)}$ and for $\hat{u}=K$ we recover the expression $2\sqrt{D_i(\beta-q)}$ (as it is shown in Fig. \ref{fig:speed}a).

On the other hand, to our knowledge there are no rigorous analytical results for traveling waves solving Eq. \eqref{eq:p}. 
Clearly an initial condition in which the function $i$ has compact support surrounded by an area with $u=K$ cannot evolve into a traveling wave connecting $(K,0)$ to $(u^*,i^*)$, as the spatial movement of $i$ is inhibited in the areas in which $u$ is at carrying capacity. As a consequence, the classical problem of a new predator or a new infection invading an established population makes no sense in this context. On the other hand, the numerical results in Fig. \ref{fig:speed}b show the existence of a traveling wave evolving from $(0,0)$ to $(u^*,i^*)$. Let us also observe the movement depends on the local density, so the speed expression $\sqrt{D_u p/2}$ is only valid when the invading front is at carrying capacity; when the front is smaller due to the infection, it also moves slower. In the case of Fig. \ref{fig:speed}b, the invading front is close enough to $K$, so that the value $\sqrt{D_u p/2}$ is still a good approximation for the speed of uninfected invasion.

Let us conclude this section by recalling the fact that all the speed wave expressions are accurate in one spatial dimension. In two spatial dimensions, the same formulas describe the asymptotic speed for the radially symmetric equation (see for example \cite[\S 13.2]{murray02}); our numerical simulations show that the formulas of this section approximate the wave speed well enough in the parameters' range of our interest.

\begin{table}[t!]
	\centering{
		\scriptsize
		\begin{tabular}{lllll}
			&\multicolumn{1}{c}{\textbf{Parameter}} & \multicolumn{1}{c}{\textbf{Description}} &  \multicolumn{1}{c}{\textbf{Value [Units]}} & \multicolumn{1}{c}{\textbf{Reference}}\\
			\hline
			\\
			& $p$         &maximal duplication rate of uninfected cells     & $1.87 \times 10^{-2}$ [h$^{-1}$] &\cite{ke00} \\
			& $q$         &death rate of infected cells     & $4.17\times 10^{-2}$ [h$^{-1}$]	&\cite{ganly00} \\
			& $D_u, D_i$         &diffusion coefficients (undirected movement)    & $1.88\times 10^{-4}$ [mm$^2$/h]	&estimate based on \cite{kim06} \\
			& $D_u, D_i$         &diffusion coefficients (pressure-driven movement)    & $1.50\times 10^{-3}$ [mm$^2$/h]	&estimate based on \cite{kim06} \\
			& $K^{1 \text{D}}$         &tissue carrying capacity in one dimension   & $10^3$ [cells/mm]	&model estimate \\
			& $K^{2 \text{D}}$         &tissue carrying capacity in two dimensions   & $10^4$ [cells/mm$^2$]	&\cite{lodish08} \\
			& $\beta$         &infection rate   & $1.02\times 10^{-1}$ [h$^{-1}$]	&estimate based on \cite{friedman06} \\
			& $R_u$         &initial radius of uninfected cells     & $2.6$ [mm]	&\cite{kim06} \\
			& $R_i$         &initial radius of infected cells     & $1$ [mm]	&model estimate \\
			\\
			\hline
		\end{tabular}
	\caption{Reference parameter set.}\label{tab:parameters}}
\end{table}
	
\section{Comparison of the models with undirected movement}
\label{sec:undirected}

\begin{figure}
	\centering
	\includegraphics[width=0.9\linewidth]{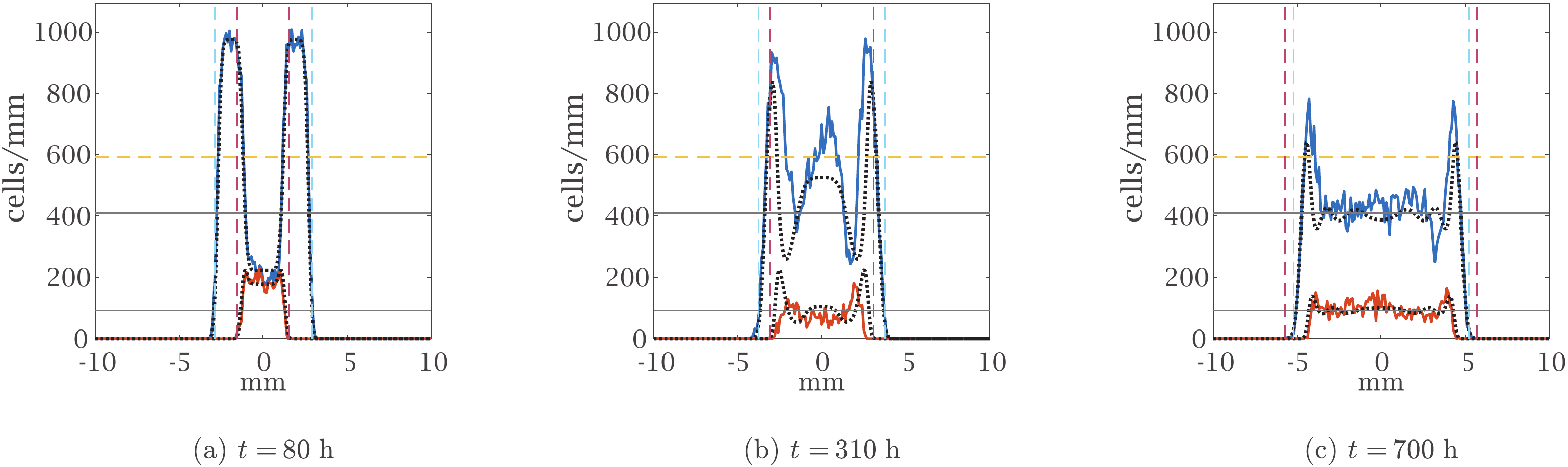}
	\caption{Comparison in one spatial dimension between numerical simulations of the discrete model with undirected movement (solid lines) and the numerical solution of Eq. \eqref{eq:l} (dotted black lines) at three different times, with the parameters given in Table \ref{tab:parameters}. For the agent-based model, the density of the uninfected cells is represented in blue and the density of infected cells in red. The vertical dashed lines represent the expected positions of the uninfected and infected invasion fronts, traveling respectively at speed $2\sqrt{D_u p}$ (blue lines) and $2\sqrt{D_i(\beta-q)}$ (red lines); the latter has no biological meaning in panel (c), as the infection cannot go beyond the uninfected front. The horizontal solid black lines show the equilibrium of the ODE given by Eq. \eqref{eq:eq_l} and the horizontal dashed yellow line represents the expected uninfected density at the front given by Eq. \eqref{eq:ubar} (only relevant in panel (c)). The results of the agent based model are averaged over five simulations and the maximum of the cell density axis corresponds to the maximum over time of this average (which is lager than the carrying capacity).}
	\label{fig:l_oned}
\end{figure}

\begin{figure}
	\centering
	\includegraphics[width=0.9\linewidth]{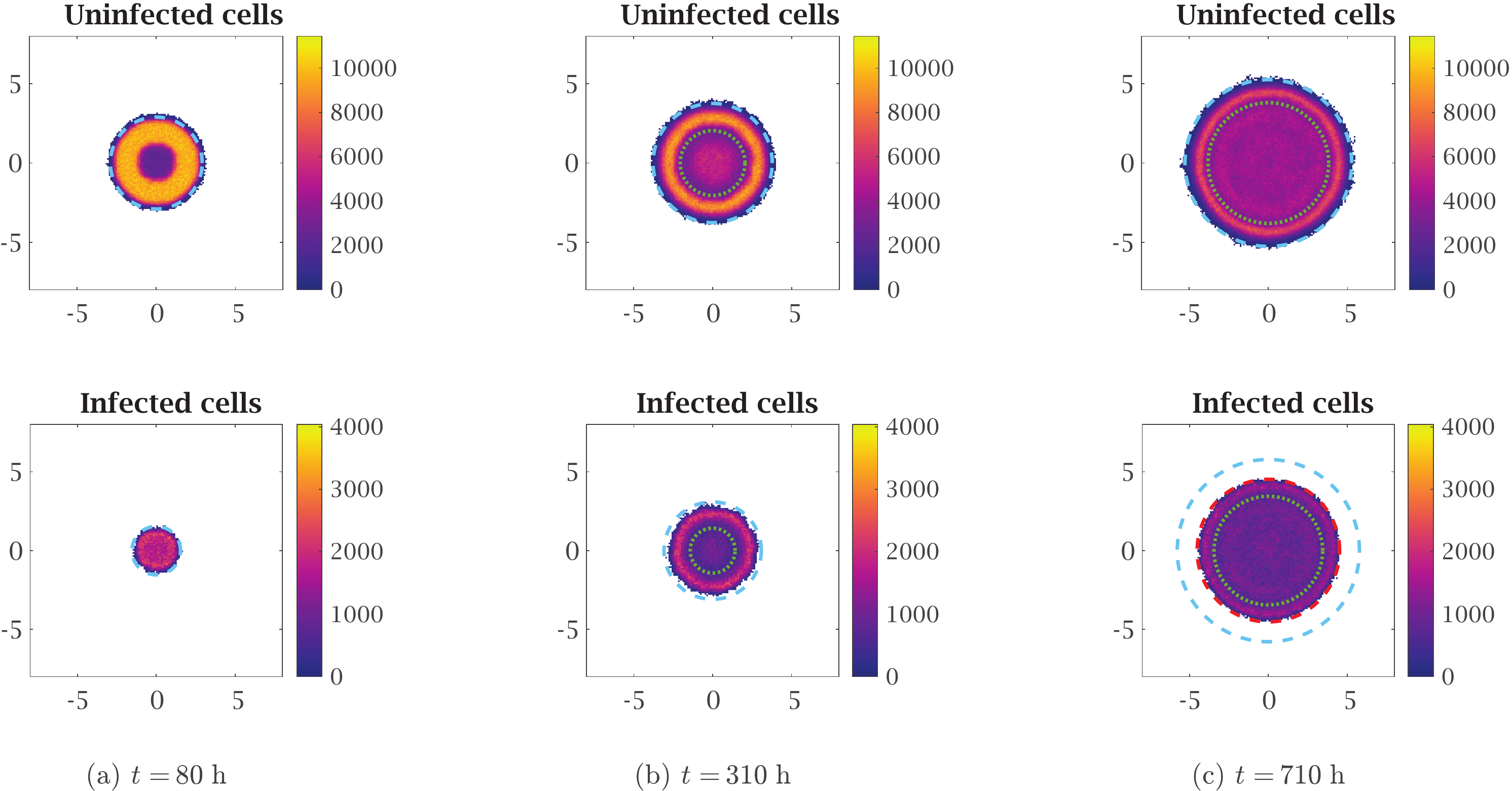}
	\caption{Numerical simulation of the discrete model with undirected movement in two spatial dimensions at three different times with the parameters given in Table \ref{tab:parameters}. The dotted green circles represent the internal minimum of the numerical solution of Eq. \eqref{eq:l} (not shown in panel (c), as this minimum is in 0). The dashed cyan circles represent the expected positions of the uninfected and infected invasion fronts, traveling respectively at speed $2\sqrt{D_u p}$ and $2\sqrt{D_i(\beta-q)}$. The latter has no biological meaning in panel (c), as the infection cannot go beyond the uninfected front; therefore in this figure we show with a dashed red circle the front of the infected cells given by the numerical solution of Eq. \eqref{eq:l}. The results of the agent based model are averaged over five simulations and the maximum of the colorbars for uninfected and infected cells correspond to the maximum over time of the averages (which for the uninfected cells is lager than the carrying capacity).}
	\label{fig:l_twod}
\end{figure}

We are now ready to compare numerical simulations for the agent-based model and the corresponding system of PDEs. We start from the model with standard diffusion, since in this case there exist comprehensive analytical results for traveling waves.
We consider a spatial domain $[-L,L]$ (or $[-L,L]^2$) with $L=10\;$mm and we adopt Neumann boundary conditions. The initial conditions are
\begin{equation}
	\label{eq:initial}
	u_0(x)=
	\begin{cases}
		0.9\ K \quad &\text{for } \abs{x}\leq R_u \\
		0\ \quad &\text{for } \abs{x}> R_u 
	\end{cases}
	\qquad
	i_0(x)=
	\begin{cases}
		0.1\ K \quad &\text{for } \abs{x}\leq R_i \\
		0 \quad &\text{for } \abs{x}> R_i 
	\end{cases}
\end{equation}
where $R_u$ and $R_i$ are respectively the initial radius of uninfected and infected cells. This corresponds to a central viral injection; a short discussion about how initial conditions affect the dynamics can be found in Appendix \ref{app:num}.

We first present the results obtained with the reference parameters listed in Table \ref{tab:parameters} in both one and two dimensions; we then investigate how different parameters allow to obtain other spatial patterns.

\paragraph{Reference parameters} 
Fig. \ref{fig:l_oned}, along with the video accompanying it (see electronic supplementary material S2), shows an excellent quantitative agreement between numerical solutions of the system of PDEs \eqref{eq:l} and the average over 5 numerical simulations of the agent-based model in one spatial dimension. At the beginning of the simulations, the central region of the tumour is quickly infected, while the outer region (which is only occupied by uninfected cells) grows up until reaching the carrying capacity. At the same time, a traveling wave of uninfected cells starts to invade the surrounding area at the speed $2\sqrt{D_u p}$ (vertical blue lines in Fig. \ref{fig:l_oned}), as predicted by theoretical results. As soon as the uninfected cells reach the carrying capacity, the invasion speed of the infected cells stabilises to the value $2\sqrt{D_i(\beta-q)}$ (vertical red lines in Fig. \ref{fig:l_oned}), which again confirms our expectations from analytical results. In the meantime, cell densities at the center of the tumour converge with damped oscillations to the equilibrium of the corresponding ODE (horizontal solid black lines in Fig. \ref{fig:l_oned}), given by Eq. \eqref{eq:eq_l}. This is shown in Fig. \ref{fig:l_oned}a.

The parameters we chose are such that $2\sqrt{D_u p}<2\sqrt{D_i(\beta-q)}$, meaning that the infection eventually reaches the front of the wave of uninfected cells. This happens around time $t=200\;$h: as a consequence, the peak at the front starts to decrease for both populations and infected cells slow down (see Fig. \ref{fig:l_oned}b). The final peak of the uninfected cells is approximately 
\begin{equation}
	\label{eq:ubar}
	\bar{u}\coloneqq \Bigl( \frac{q}{\beta}+\frac{D_u p}{D_i \beta} \Bigr) K= u^* +\frac{D_u p K}{D_i \beta}
\end{equation}
which is the solution of the equation
\[
2\sqrt{D_u p}=2\sqrt{D_i\Bigl(\frac{\beta}{K}\bar{u}-q\Bigr)}
\]
In other words, an uninfected population of cell density $\bar{u}$ is invaded by infected cells at speed $2\sqrt{D_u p}$, which is the speed of the uninfected front. A higher uninfected density at the front would result in a faster invasion of the infection, which would cause the front to decrease again; similarly, a smaller uninfected density at the front would slow down the infection and thus allow the uninfected front to grow. In our case the density of the uninfected population is not constant, but the value $\bar{u}$ given in Eq. \eqref{eq:ubar} is still a good approximations of the density at the front (see the horizontal dashed yellow line in Fig. \ref{fig:l_oned}c).
As time passes, both front waves keep moving at the speed $2\sqrt{D_u p}$; the fronts are followed by a few damped oscillations that converge to the equilibrium of the ODE. This is shown in Fig. \ref{fig:l_oned}c.

Fig. \ref{fig:l_twod}, along with the video accompanying it (see electronic supplementary material S3), shows that the same excellent agreement also holds in two spatial dimensions; the comparison with the continuum model and the analytical expressions of the wave speeds is shown through dashed and dotted colored circles, as explained in the caption of the figure. Observe that, before cell densities converge to the equilibrium in the center of the tumour, some concentric circles appear, in line with experimental observation \cite{wodarz12}; the internal circle however disappears as time passes.

\begin{figure}
	\centering
	\includegraphics[width=0.9\linewidth]{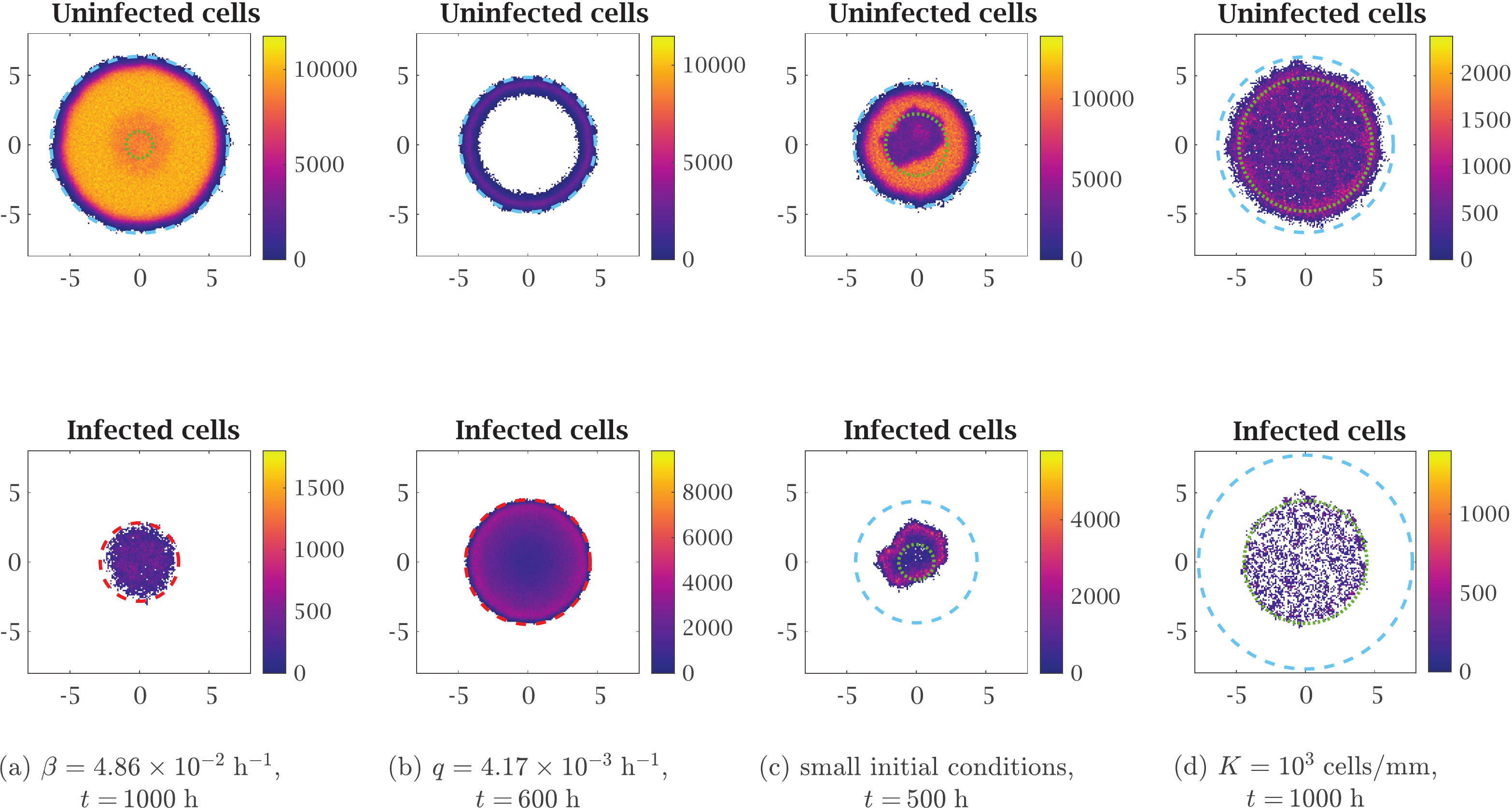}
	\caption{Numerical simulation of the discrete model with undirected movement in two spatial dimensions with different parameter values. The dotted green circles represent the internal minimum of the numerical solution of Eq. \eqref{eq:l} (not shown when this minimum is in 0). The dashed cyan circles represent the expected positions of the uninfected invasion fronts, traveling at speed $2\sqrt{D_u p}$. The dashed red circles represent the front of the infected cells given by the numerical solution of Eq. \eqref{eq:l}. The parameters employed are the ones given in Table \ref{tab:parameters}, with the exception of the infection rate $\beta$ in panel (a) (which is set to $4.86\times 10^{-2}\;$h$^{-1}$, i.e. less than half of the reference value), the death rate of infected cells $q$ in panel (b) (which is set to $4.17\times 10^{-3}\;$h$^{-1}$, i.e. one tenth of the reference value), the initial conditions in panel (c) (whose densities are set to $0.09\,K$ for uninfected cells and $0.01\,K$ for infected cells, i.e. one tenth of the reference values of Eq. \eqref{eq:initial}) and the carrying capacity $K$ in panel (d) (which is set to $10^3\;$cells/mm; initial conditions are scaled accordingly). The first two figures are the averages over five simulations, while the last two represent single simulations. In both cases the maximum of the colorbars for uninfected and infected cells correspond to the maximum over time of the quantity plotted, which for uninfected cells is lager than the carrying capacity (note the different values between different simulations).}
	\label{fig:l_varie}
\end{figure}

\paragraph{Impact of the parameters on the treatment outcome}
Let us show how varying the parameters affects the success of the therapy, still for the case of growth that is unhindered by spatial or pressure constraints, but is only limited by carrying capacity. We only focus on two-dimensional simulations, but the one-dimensional case is analogous.

We start by analysing some instances of treatment failure. As we already pointed out in Section \ref{sec:wave}, the worst possible case is the situation in which the infection ceases after a finite time and uninfected cells grow at carrying capacity: this corresponds to parameter values such that $\beta<q$, which do not allow the equilibrium $(u^*,i^*)$ to be positive. A more interesting case of failure, which has no analogue in the spatially homogeneous ODE, is the one in which the equilibrium $(u^*,i^*)$ is positive and stable, the infected cells form a traveling wave, but the spread of the infection is smaller than the speed of the uninfected wave and so the outer region of the tumour is completely unaffected by the therapy. Fig. \ref{fig:l_varie}a shows this situation, obtained by decreasing the infection rate $\beta$ with respect to the reference value. In this case the value of $u^*$ is more than 85\% of the carrying capacity, so the invasion front is at carrying capacity and even in the central area the role of the infection is not really relevant, despite never ceasing completely. Similar situations are obtained whenever parameter values are such that
\[
2\sqrt{D_i(\beta-q)}<2\sqrt{D_u p}
\]
If $D_u=D_i$, this condition is equivalent to $\beta<q+p$. We can thus conclude that, as we could easily expect, a decrease in the infection rate $\beta$ or an increase either of the death rate of the infected cells $q$ or the proliferation rate of the uninfected cells $p$ with respect to the reference value makes the therapy less successful and, in extreme cases, useless. This scenario mimics, to some extent, that of an aggressively expanding tumour whose developing front is moving very fast, as in existing clinical settings~\cite{Eissa2018review}.

On the other hand, whenever the infection reaches the boundary of the tumour (as in the reference situation) we can consider the therapy at least partially successful. Some variations of the parameter values allow then to improve therapy achievements. For example, as the death rate of the infected cells $q$ decreases, the infection propagates faster and $u^*$ decreases, therefore the therapy becomes more effective. This situation is shown in Fig. \ref{fig:l_varie}b and captures the typical case when the virus has sufficient potency, as current clinical trials and therapeutic practice strive to achieve~\cite{lawler17,Hemminki2020review}. The center of the tumour is almost completely void for most of the time, as the number of uninfected cells is negligible and the number of infected cells is quite small (although slightly bigger). At later times some other inner circles emerge as a consequence of the damped oscillations leading to the equilibrium; nevertheless, the emerging spatial structure can still be well described as an empty ring. It is clear from these results that a way to make the therapy more efficient would be to increase $\beta$, as this would again result in a faster infection and a smaller uninfected population. A decrease of $p$ would leave the number of uninfected cells at the equilibrium unchanged; yet, the tumour expansion would slow down and, as a consequence, the infection would reach the tumour boundary faster. Unlike the continuous model, the agent-based model may show extinction in finite time of both populations, which correspond to the eradication of the tumour (not shown here). However, this would require to change parameters beyond the values that appear biologically meaningful. This is in line with results obtained from deterministic spatially homogenous models, for example the simple one in \cite{jenner18biomath}.

Table \ref{tab:waves} summarises the different scenarios for traveling waves described above. In all these cases the results of the agent-based model perfectly agree with the ones given by the numerical solution of the corresponding PDE. Let us stress the fact that taking into account a single simulation in most of the cases reduces the quantitative agreement, but not the overall qualitative behavior: individual variations occur but a general, consistent trend is achieved.

\begin{table}[t!]
	\centering{
		\scriptsize
		\begin{tabular}{ccp{0.6\linewidth}}
			&\multicolumn{1}{c}{\textbf{Parameter}} & \multicolumn{1}{c}{\textbf{Description}}\\
			\hline
			\\
			& $\beta<q$     & uninfected cell wave at speed $2\sqrt{D_u p}$ and height $K$, no infection.\\
			\hline
			\\
			& $q<\beta<q+\frac{D_u}{D_i}p$   & uninfected cell wave at speed $2\sqrt{D_u p}$ and height $K$, central infection expanding at speed $2\sqrt{D_i (\beta-q)}$ without reaching the uninfected front; internal densities reach the values $\Bigl( \frac{qK}{\beta}, \frac{pK(\beta -q)}{\beta(\beta +p)}\Bigr)$.\\
			\hline
			\\
			& $\beta>q+\frac{D_u}{D_i}p$   & uninfected cell wave at speed $2\sqrt{D_u p}$ and height $\Bigl( \frac{q}{\beta}+\frac{D_u p}{D_i \beta}\Bigr) K$, infection up to the uninfected front; internal densities reach the values $\Bigl( \frac{qK}{\beta}, \frac{pK(\beta -q)}{\beta(\beta +p)}\Bigr)$.\\
			\\
			\hline
		\end{tabular}
		\caption{Summary of the different scenarios for the traveling waves as the infection rate $\beta$ increases.}\label{tab:waves}}
\end{table}

\paragraph{Impact of stochasticity for lower cell densities}
We now present two simulations in which stochastic effects give rise to notable differences between the discrete and continuum approach, due to the fact that a smaller number of cells reduces the quality of the continuum approximation. This could correspond to a moderately extended tumour in its first stages of growth, for example. Fig.  \ref{fig:l_varie}c shows the result of a single simulation with the parameters of Table \ref{tab:parameters} and smaller initial cell densities. Clearly, uninfected cells take longer than in the reference case to reach carrying capacity. As soon as they do, the infected area is much less regular than what the PDE predicts: this comes from the fact that the infection starts among a small number of cells and thus a few stochastic events affect the spatial distribution of the infection relevantly. As times passes, these differences tend to disappear.

Fig.  \ref{fig:l_varie}d shows the situation in which the carrying capacity $K$ is decreased and initial cell densities are scaled accordingly, in agreement with Eq. \eqref{eq:initial}. For initial times it is still possible to recognise the same qualitative behavior of the PDEs, but as time passes stochastic events drive the system into a very irregular spatial configuration.

Let us also mention what happens when we change the reference parameters for the scaled system (not shown here). If we decrease the death rate of infected cells, we still observe the void ring structure, although much less precise than the one of Fig. \ref{fig:l_varie}b. If the infection rate is decreased as in Fig. \ref{fig:l_varie}a, then the number of infected cells is so low that the infection undergoes extinction in short time. We can therefore conclude that the PDEs remain a good description of the treatment outcome, even though quantitative agreement is lost due to stochastic effects.

\begin{figure}
	\centering
	\includegraphics[width=0.9\linewidth]{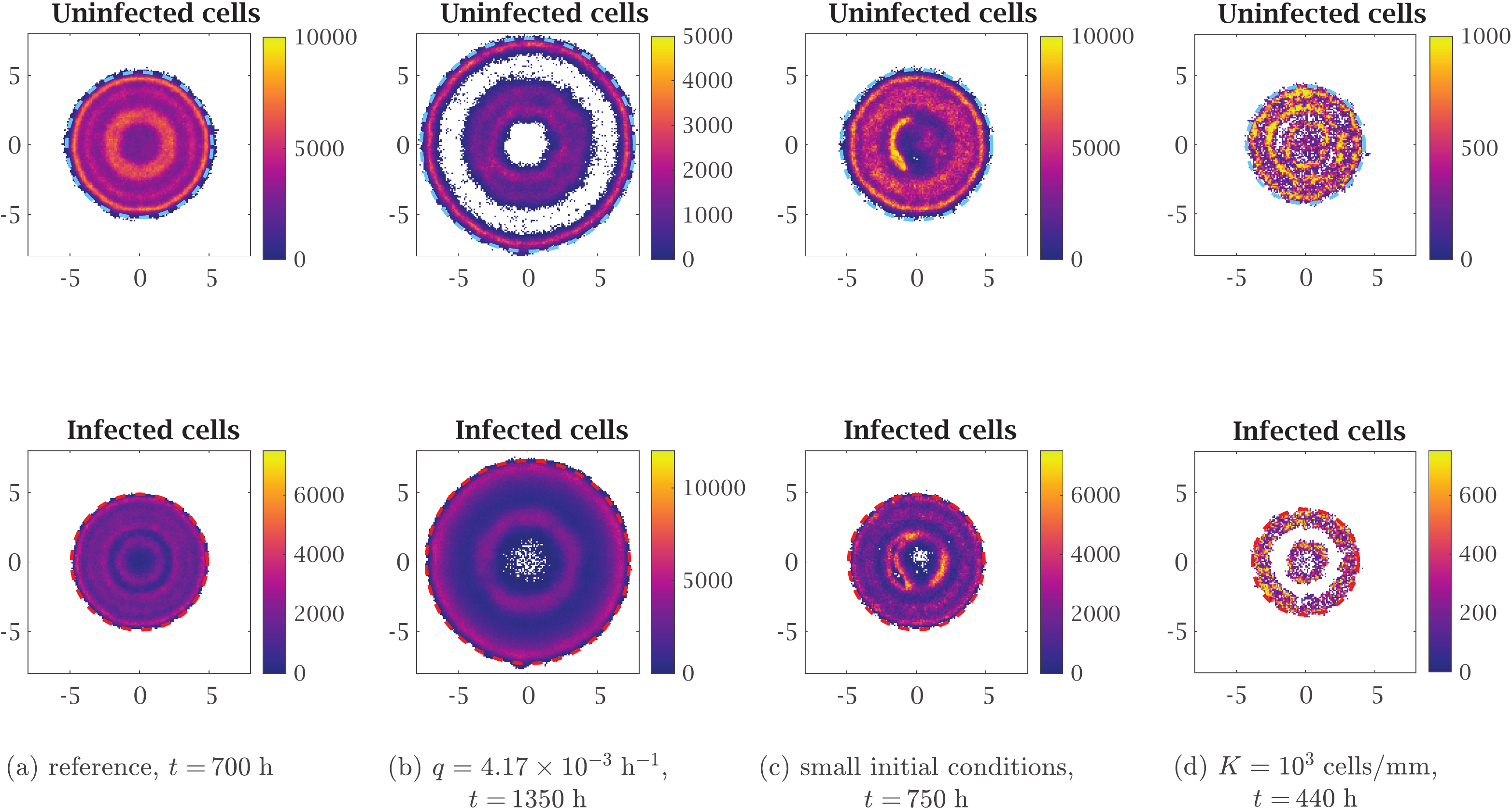}
	\caption{Numerical simulation of the discrete model with undirected movement and exponential growth in two spatial dimensions, with different parameter values. The dashed cyan circles represent the expected positions of the uninfected invasion fronts, traveling at speed $2\sqrt{D_u p}$. The dashed red circles represent the front of the infected cells given by the numerical solution of Eq. \eqref{eq:l}. The parameters employed are the ones given in Table \ref{tab:parameters}, with the exception of the death rate of infected cells $q$ in panel (b) (which is set to $4.17\times 10^{-3}\;$h$^{-1}$, i.e. one tenth of the reference values), the initial conditions in panel (c) (whose densities are set to $0.09\, K$ for uninfected cells and $0.01\, K$ for infected cells) and the carrying capacity $K$ in panel (d) (which in this case only affects the infection and is set to $10^3\;$cells/mm; initial conditions are scaled accordingly). The first two figures are the averages over five simulations, while the last two represent single simulations. The maximum values of the colorbars have been chosen in order to make the figures clear and are much smaller than the maximum reached by uninfected cell densities (note the different values between different simulations).}
	\label{fig:n_varie}
\end{figure}

\paragraph{Exponential growth} One may wonder whether the growth of a small tumour that is not limited by the lack of external resources may be stopped only by viral infections. Unlimited exponential growth is clearly not feasible in any biological scenario, but we could imagine that in some cases the carrying capacity is too high to give any significant contribution in the initial phases of the tumour dynamics. We thus let $G(\rho)\equiv p$ and study what happens in the situations we have analysed so far.

The internal equilibrium of the associate ODE is $(qK/\beta,pK/\beta)$ and it is neutrally stable. It still makes sense to look for traveling waves connecting this two equilibria, as the addition of diffusion to the system is enough to make the equilibrium asymptotically stable; it is also reasonable to expect the oscillations to take longer to dampen than in the previous situation because of this. Nevertheless, to our knowledge there are no rigorous analytical results for traveling waves solutions of the associate PDE.

Fig. \ref{fig:n_varie}a, along with the video accompanying it (see electronic supplementary material S4), shows the result of the two-dimensional simulation with the parameters of Table \ref{tab:parameters}. The supplementary material shows that also in the present case there is an excellent agreement between the discrete and continuum model. At the beginning of the simulation, uninfected cells in the outer region grow exponentially and invade the surroundings at speed  $2\sqrt{D_u p}$ (dashed cyan circles in Fig. \ref{fig:n_varie}a). Meanwhile, the speed of infected cells increases as the number of uninfected cells grows, until the infection eventually reaches the front of uninfected cells: this happens around time $t=100\;$h, which is approximately half the time it takes for the same process with logistic growth; however, in this case the peak of uninfected cells is more than five times $K$. After that, the peak of uninfected cells quickly drops to approximately the value $\bar{u}$ predicted by Eq. \eqref{eq:ubar}. In the center of the tumour there are several secondary waves with a peak of size comparable to the front peak, which propagate both toward the interior and the exterior of the tumour. These waves are lead by uninfected cells, with infected cells following: when two uninfected waves merge, they quickly disappear because they get surrounded by infected cells. 

The case of an ineffective treatment does not exist mathematically, as the equilibrium values $(qK/\beta,pK/\beta)$ are positive for all values of the parameter. Furthermore, the propagation speed of the infection increases as the number of infected cells increase: since the growth is unlimited, the infection eventually and inevitably reaches the front of the uninfected cells. The shortcoming is the fact that all the dynamics happen at much higher density levels than those considered previously, and, as such, appear biologically irrelevant. On the other hand, the situation of a highly effective therapy does not present any relevant difference with respect to the situation with logistic growth: Fig. \ref{fig:n_varie}b shows that a decrease of the death rate of infected cells yields a result very similar to the one obtained in Fig.  \ref{fig:l_varie}b: the only difference is that inner circles are more visible and persistent at late times.

Let us also mention what happens for low cell densities: we expect stochasticity to play a more important role in this model, as there is no deterministic limit to cell growth. Fig. \ref{fig:n_varie}c shows that scaling only the initial conditions leads again to spatial patterns that are much less regular than what the PDEs predicts; furthermore, these features are still evident even for long times. Fig.  \ref{fig:n_varie}d shows that as we scale the whole system by a factor of ten we still maintain some qualitative agreement with the PDEs at early times, but the importance of stochastic effects becomes evident; we also observe that local peaks may get very high before the infection manages to control them. Despite the increasing importance of stochasticity in this situation, it is important to observe that the PDEs are still able to correctly predict the outcome of the therapy as parameters change.

\section{Comparison of the models with pressure-driven movement}
\label{sec:pressure}

\begin{figure}
	\centering
	\includegraphics[width=0.9\linewidth]{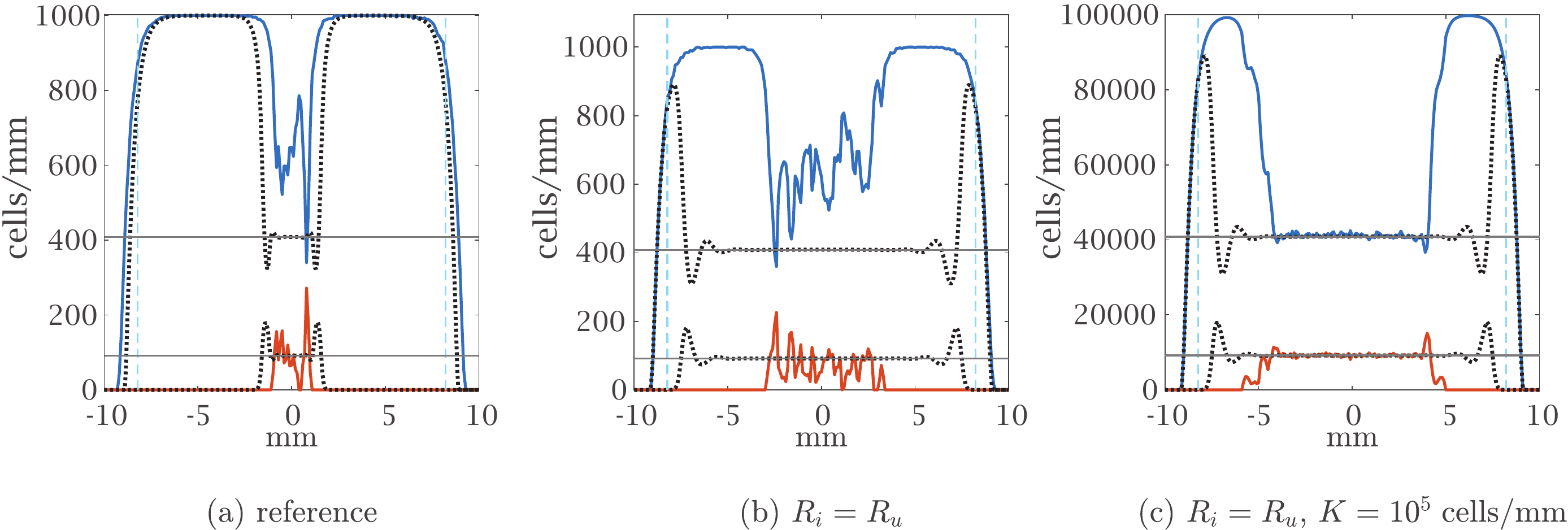}
	\caption{Comparison in one spatial dimension between numerical simulation of the discrete model with pressure-driven movement (solid lines) and the numerical solution of Eq. \eqref{eq:p} (dotted black lines) at the same time $t=1500\;$h with different parameter values, all resulting in treatment failure according to the discrete model. For the agent-based model, the density of the uninfected cells is represented in blue and the density of infected cells in red. The vertical dashed blue lines represent the expected positions of the uninfected invasion front, traveling at speed $\sqrt{D_u p/2}$. The horizontal solid black lines show the equilibrium of the ODE given by Eq. \eqref{eq:eq_l}. The parameters employed are the ones given in Table \ref{tab:parameters}, with the exception of the infection radius $R_i$ in panels (b) and (c) (which is set to $2.6\;$mm) and the carrying capacity $K$ in panel (c) (which is set to $10^5\;$cells/mm).
		The results of the agent based model are averaged over five simulations and the maximum of the cell density axis corresponds to the maximum of this average.}
	\label{fig:p_oned_fail}
\end{figure}

\begin{figure}
	\centering
	\includegraphics[width=0.9\linewidth]{Fig8}
	\caption{Numerical simulation of the discrete model with undirected movement in two spatial dimensions at the same time $t=1500\;$h with different parameter values, all resulting in treatment failure. The dotted green circles represent the internal minimum of the numerical solution of Eq. \eqref{eq:p} (not shown when this minimum is in 0). The dashed cyan circles represent the expected positions of the uninfected invasion fronts, traveling at speed $\sqrt{D_u p/2}$. The dashed red circles represent the front of the infected cells given by the numerical solution of Eq. \eqref{eq:p}. The parameters employed are the ones given in Table \ref{tab:parameters}, with the exception of the infection radius $R_i$ in panels (b) and (c) (which is set to $2.6\;$mm) and the carrying capacity $K$ in panel (c) (which is set to $10^5\;$cells/mm$^2$). The results of the agent based model are averaged over five simulations and the maximum of the colorbars for uninfected and infected cells correspond to the maximum over time of the averages. Note the finger-like formations.}
	\label{fig:p_twod_fail}
\end{figure}

Finally, let us discuss the numerical simulations for the model with pressure-driven movement and logistic growth. As we already pointed out, the linear spreading speed does not give any meaningful information. An additional difficulty comes from the fact that varying initial conditions may result in opposite therapy outcomes: this is a consequence of cells' inability to propagate in areas of constant total density. We will also see that the role of stochasticity is more important than in the previous models and in many cases the PDEs are unable to correctly predict the therapy outcome. This represents an important insight when assessing the efficiency of virotherapy for tumours that are either highly constrained or are hard to infect or penetrate.

Given the intrinsic variability, we do not give a comprehensive description of all the possible outcomes in the way we did in the previous section and limit to the description of some cases of failure and success of the therapy, with a special emphasis on the situations in which results from agent-based model and PDE do not agree.

We again adopt the initial conditions given by Eq. \eqref{eq:initial} and Neumann boundary conditions.

\paragraph{Reference parameters: ineffective treatment}
Let us first analyse Figs. \ref{fig:p_oned_fail}a and \ref{fig:p_twod_fail}a, which show an excellent quantitative agreement between numerical solutions of the system of PDEs \eqref{eq:p} and the average over five numerical simulations of the agent-based model both in one and two spatial dimensions. Unlike the previous situations, this model predicts the infection to be confined at the center of the tumour: this is due to the fact that the central infection quickly causes the total cell density to drop, while external uninfected cells proliferate; since cells cannot move toward an area with higher cell density, the outer cells are never going to be infected and the tumour keeps expanding at the speed  $\sqrt{D_u p/2}$ (vertical dashed blue line in \ref{fig:p_oned_fail}a and dashed cyan circle in Fig. \ref{fig:p_twod_fail}a), in the same way it would do in absence of treatment. This situation is similar to the case of ineffective infection already observed in Fig. \ref{fig:l_varie}a, but it is important to observe that here the infection rate has not been decreased with respect to the reference value. Therefore, it is clear that in this model constraints to cell movement are responsible for treatment failure. 

It is important to remark that adding explicit viral dynamics to the model and allowing the virus to diffuse without any constraint due to crowding effects (as, for example, in \cite{pooladvand21}) would result in an effective infection even in the case of pressure-driven cell movement, but does not entirely capture the realism of the process. We are considering a situation in which the virus faces some challenges in penetrating the tumour and thus cell movement is clearly a mayor driver of viral propagation.

\paragraph{Treatment success in the continuous setting}
Since treatment failure is due to the inability of the infection to propagate in the tumour, a simple solution to improve outcomes could be to consider that infected cells are initially present in the whole tumour, i.e. take $R_i=R_u$ in Eq. \eqref{eq:initial}. From the biological point of view, this corresponds to multiple locations for the initial viral injection in contrast to a single central injection, which has been considered so far. Figs. \ref{fig:p_oned_fail}b and \ref{fig:p_twod_fail}b indeed show that using this approach the PDEs predict infected cells to be at all times at the tumour front, giving rise to traveling waves qualitatively similar to the ones we observed in the model with undirected movement. Nevertheless, the agent-based model again shows an infection that fails to propagate in the whole tumour. This is due to the fact that, in this model, demographic stochasticity plays a much more important role than in previous models: any growth above average of uninfected cells stops the movement of infected cells and hence cannot be compensated at later times by other processes. Let us also observe that the PDE predicts the presence of a very small infected cell density up to the uninfected invasion front: in the discrete model, this corresponds to a number of infected cells too low to guarantee a good quality of the continuous approximation (in these regards, see also the discussions in \cite{johnston20,macfarlane22}). Overall, at these scale the discrete model cannot be accurately described by the continuum model.

According to the formal derivation of the PDEs from the agent-based model, an increase of cell number and a decrease of the temporal and spatial discretisation improves the quality of the continuum approximation. Hence, we scaled the system by setting $K=10^5\;$cells/mm in one dimension and $K=10^5\;$cells/mm$^2$ in two dimensions. While this increase has no biological justification, from the mathematical point of view it still makes sense to analyse at what scale we obtain good agreements between the discrete and the continuous model. Figs. \ref{fig:p_oned_fail}c and \ref{fig:p_twod_fail}c show that, despite an excellent quantitative agreement at initial times, stochastic events at some point inevitably cause external cells to start to grow: a positive feedback loop then promotes cellular growth until carrying capacity, stopping any further spatial propagation of the infection. We can thus conclude that only a further increase of the cell number could guarantee a better agreement between the discrete and the continuum model, although the biological meaning would be lost.

\begin{figure}
	\centering
	\includegraphics[width=0.9\linewidth]{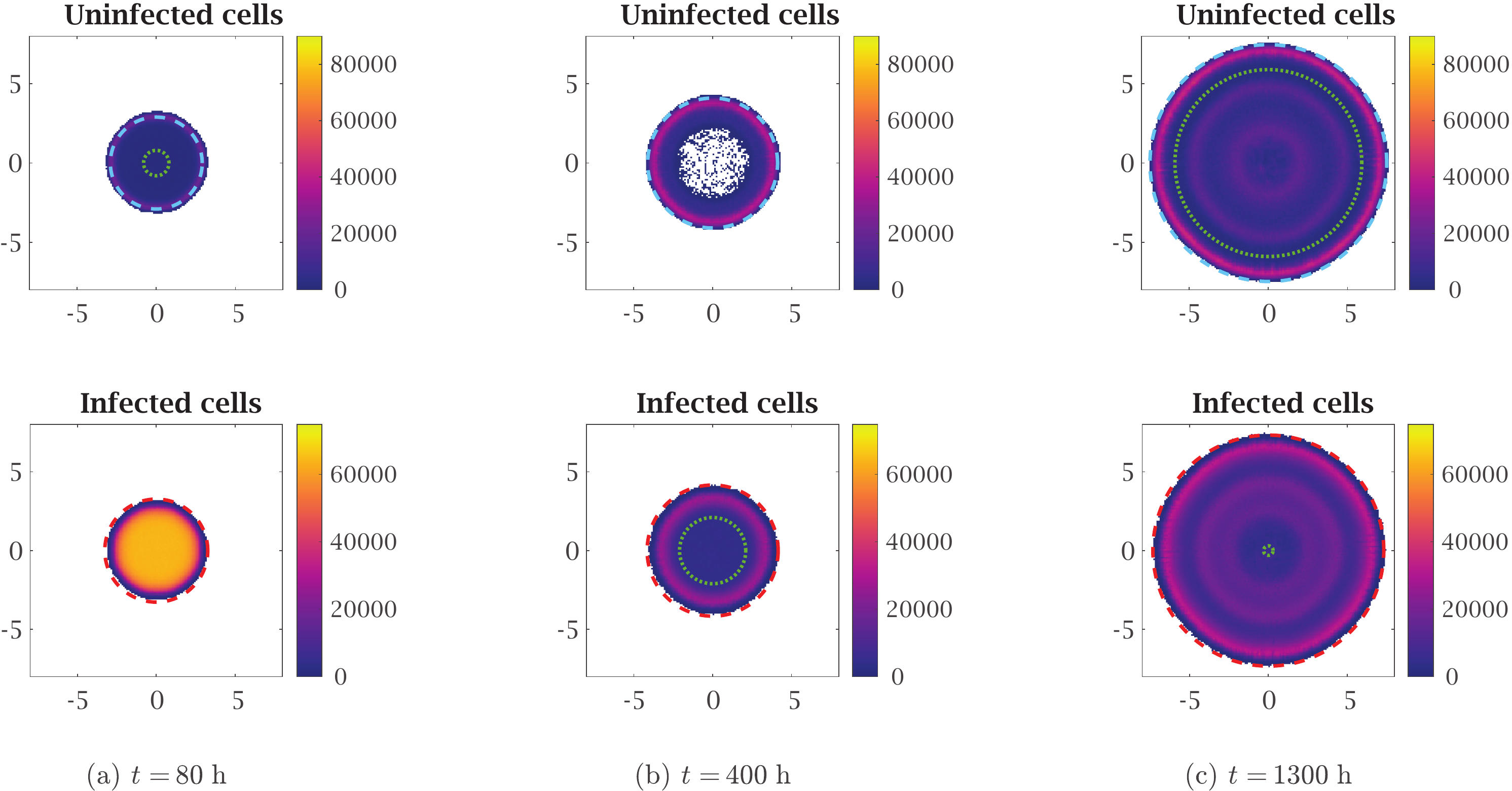}
	\caption{Numerical simulation of the discrete model with pressure-driven movement in two spatial dimensions at three different times. The dotted green circles represent the internal minimum of the numerical solution of Eq. \eqref{eq:p} (again, not shown when this minimum is in 0). The dashed cyan circles represent the expected positions of the uninfected invasion fronts in absence of treatment, traveling at speed $\sqrt{D_u p/2}$. The dashed red circles represent the front of the infected cells given by the numerical solution of Eq. \eqref{eq:p}. The parameters employed are the ones given in Table \ref{tab:parameters}, with the exception of the carrying capacity $K$ (which is set to $10^5\;$cells/mm$^2$, i.e. ten times the reference value) and the death rate of infected cells $q$ (which is set to $8.33\times 10^{-3}\;$h$^{-1}$, i.e. one fifth of the reference values). These parameter choices allow a perfect agreement between the discrete and the continuous model, although their biological value is disputable.
		The results of the agent based model are averaged over five simulations and the maximum of the colorbars for uninfected and infected cells correspond to the maximum over time of the averages.}
	\label{fig:p_twod_sucq}
\end{figure}

\paragraph{Treatment success in the discrete setting}
Let us now describe two parameter settings that allows the discrete model to create traveling wave, so that the therapy is at least partially successful. Figs. \ref{fig:p_twod_fail}b and \ref{fig:p_twod_fail}c show that in two dimensions the infection propagates more easily than in one dimension, as there is more space to overcome the unexpected growth of uninfected cells at single points; we therefore expect to observe traveling waves in two dimensions by changing the reference parameter values in ways less significant than in one dimension. Indeed, Fig. \ref{fig:p_twod_sucq}, along with the video accompanying it (see electronic supplementary material S6), shows that an increase in the number of cells and a decrease in the death rate of infected cells $q$ give rise to a wave in the two dimensional discrete model, in agreement with the numerical solution of the PDE. We recall that, in the model with undirected cell movement, the decrease of the parameter $q$ is associated with a highly effective therapeutic outcome; we thus have an additional confirmation that in the discrete model with pressure-driven movement partial success is not viable.

As we have already mentioned, in one spatial dimension a good propagation of infection in the discrete model is harder to achieve. The electronic supplementary video S5 shows that a good agreement between the agent-based model and the numerical solution of the PDE is still possible, but can only be attained in unrealistic parameter ranges. Observe that in that simulation the diffusion coefficients are much higher than the reference values, indicating again that cell movement is the main obstacle to be overcome for a full success. 

In both cases, reasonable increases of the infection rate $\beta$ do not lead to a more effective infection, as this causes a decrease of central cell density and creates the need for infected cells to move against a pressure gradient. Clearly, further increases of $\beta$ allow for a fast eradication of the tumour in the case of spread infection and the problem caused by the inhibition of movement becomes irrelevant; this however can be attained only if we go beyond the biologically meaningful setting. 

\paragraph{Other spatial patterns} In this model the role of stochasticity is so important that we can see irregular configurations even maintaining the carrying capacity at the reference value. Furthermore, let us set the death rate of infected cells at the same value of the simulation in Fig. \ref{fig:p_twod_sucq}, as otherwise the therapy would not be effective in the discrete model. 

\begin{figure}
	\centering
	\includegraphics[width=0.9\linewidth]{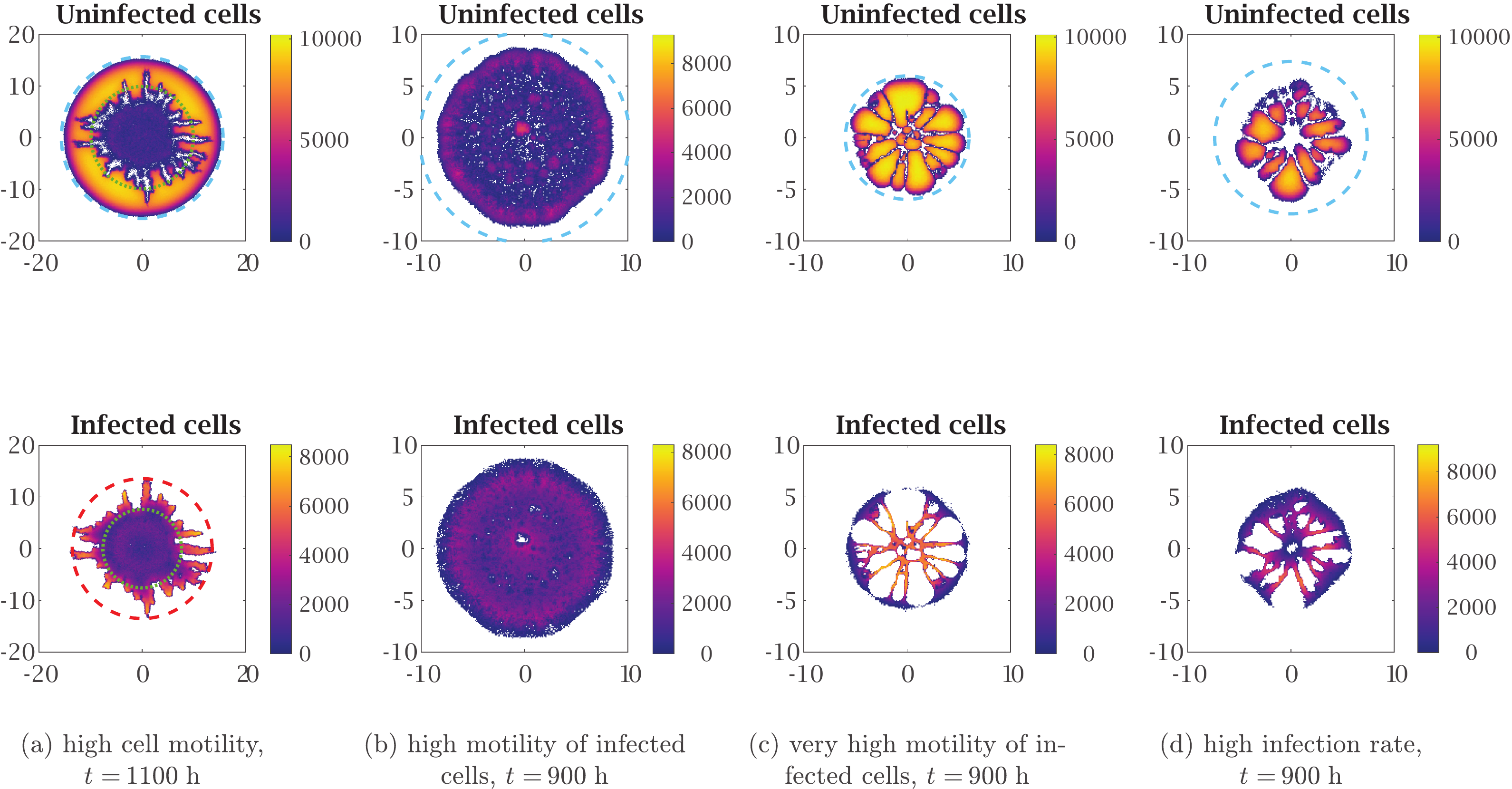}
	\caption{Numerical simulation of the discrete model with pressure-driven movement in two spatial dimensions with different parameter values. The dotted green circles in panel (a) represent the internal minimum of the numerical solution of Eq. \eqref{eq:p}. The dashed cyan circles represent the expected positions of the uninfected invasion fronts in absence of treatment, traveling at speed $\sqrt{D_u p/2}$. The dashed red circle in panel (a) represents the front of the infected cells given by the numerical solution of Eq. \eqref{eq:p}.
		The parameters employed are the ones given in Table \ref{tab:parameters}, with the exception of the death rate of infected cells $q$ (which is set to $8.33\times 10^{-3}\;$h$^{-1}$, i.e. one fifth of the reference values), the diffusion coefficient of infected cells $D_i$ (which is set to $1.50\times 10^{-1}\;$mm$^2$/h, i.e. ten times the reference value) the diffusion coefficient of uninfected cells $D_u$ in panels (a), (b) and (d) (which is set to $1.50\times 10^{-1}\;$mm$^2$/h in panel (a), to  $7.50\times 10^{-2}\;$mm$^2$/h in panel (b) and to $3.00\times 10^{-3}\;$mm$^2$/h in panel (d)) and the infection rate $\beta$ in panel (d) (which is set to $2.04\times 10^{-1}\;$h$^{-1}$, i.e. twice the reference value).
		The maximum of the colorbars for uninfected and infected cells correspond to the maximum over time of the simulation.}
	\label{fig:p_varie}
\end{figure}

Fig.  \ref{fig:p_varie}a shows that in this settings an increase of  diffusion coefficients allows the infection to propagate until approximately $1000\;$h, when it starts to be blocked by the increase of the uninfected front. Observe how the stochastic events stopping the infection take place at different times in different locations, giving rise to interesting finger shaped structures. The other simulations depicted in Fig. \ref{fig:p_varie} have been obtained by considering a higher probability of movement of infected cells with respect to uninfected cells: while there is no clear biological evidence supporting this assumption, we may still interpret it as a way to indirectly incorporate in our model, for example, a viral diffusion that is slightly more efficient in the tumour microenvironment (so that both cell-to-cell contacts and free viral particles contribute to new infections and thus the therapy is only partially inhibited by the pressure). Fig. \ref{fig:p_varie}b shows that the lower motility of uninfected cells allows the infection to occupy the whole tumour area. In a few areas uninfected cells manage to survive and become harder to be infected as they keep growing, but the therapy can still be considered effective. A further decrease of uninfected motility does not improve the situation: Fig. \ref{fig:p_varie}c shows that, despite a very effective initial infection, a few cells manage to survive and give rise to segregated structures that are almost impossible to infect, due to the low uninfected cell motility. In the majority of the tumour, uninfected cells are at carrying capacity and the tumour invasion of the surrounding tissues has been only slightly slowed down with respect of the case without infection. We can thus conclude that such a high difference in the motilities does not favor the therapy. Finally, let us consider again the value of $D_u$ used for Fig. \ref{fig:p_varie}b and double the infection rate $\beta$: as we may expect, this kind of infection makes the pressure decrease in the infected areas and it is thus too fast to be effective. Fig. \ref{fig:p_varie}d shows the result of this simulation, which is much more similar to \ref{fig:p_varie}c than to \ref{fig:p_varie}a. It is interesting to observe that this strong segregation happens with parameter values quite close to the ones that would cause a highly effective treatment, indicating how delicate the balance between the different populations is. The general message is that a pressure-driven scenario generates patterns and structures that can be hard for the virus to clear.

\section{Conclusions}
\label{sec:conclusion}

A minimal, individual-based model for the infection of tumour cells due to oncolytic viruses, assuming two different mechanisms for cellular movement, has been developed. In both cases we formally derive the deterministic continuum counterpart and compare the numerical results in one and two spatial dimensions. The outcomes of the comparison are highly dependent on the rules governing cells' movement and show typical traits for failure and successful outcomes.

In the model with undirected cell movement the solution of Eq. \eqref{eq:l} faithfully mirrors the qualitative and quantitative properties of the results of the simulations of the agent-based model: this agreement is robust to parameter variations and holds even if the logistic growth is replaced by an exponential growth. When lower cell densities are considered, the quantitative agreement is partially lost, but the PDEs are still able to correctly predict the treatment outcome. We can thus use our knowledge of the continuous model to better understand the outcome of the therapy in different parameter regimes and establish strategies and trends to help clinicians.

On the other hand, in the model with pressure-driven cell movement the solution of Eq. \eqref{eq:p} exhibits traveling waves in situations in which simulations of the agent-based model result in a localised infection in the center of the tumour, especially in one dimension. From the mathematical point of view, this can be addressed by increasing the number of agents in the simulations and decreasing the temporal and spatial discretisations. However, from the biological point of view it makes no sense to consider such a high cell density and stochastic effects cannot simply be neglected: therapy may fail only because of the inhibition of movement due to the pressure. This represents quite a hurdle from the treatment's perspective and suggests that, in the absence of an immune response, virotherapy is intrinsically limited for tumours whose microenvironments constrain cell movement.

Note also that the two dimensional patterns obtained from the agent-based simulations are consistent with the ones discussed in the literature regarding oncolytic viral infection: for example, in \cite{wodarz12} the authors describe filled rings (similar to our Figs.  \ref{fig:l_twod} and \ref{fig:l_varie}d), hollow rings (similar to our Fig.  \ref{fig:l_varie}b), concentric rings (similar to our Figs.  \ref{fig:n_varie}, \ref{fig:p_twod_sucq}) and disperse patterns (similar to our Figs. \ref{fig:l_varie}d, \ref{fig:n_varie}d, \ref{fig:p_varie}b) obtained both via in silico experiments and numerical simulations of an agent-based model. But, in \cite{wodarz12} only a single cell can occupy a lattice point and therefore concentric rings are due to stochasticity, whereas they are originated also by PDEs in our model. Results are also consistent with the spatial patterns observed in \cite{kim14}, for glioma and ECM-degrading enzyme Chase-ABC. Structures like these appear to be universal whenever tumour expansion is hindered.

We were also able to obtain segregated regions of uninfected cells (Figs. \ref{fig:p_varie}c and \ref{fig:p_varie}d) by considering a faster movement for infected cells in the agent-based model when diffusion is pressure-driven. This kind of results resemble those of stochastic invasion models \cite{lewis00,lewispacala00} and deterministic PDEs of predator and prey with an Allee effect, due to the instability of the propagation front \cite{li15,morozov06,petrovskii05,petrovskii02}. Unlike these two models though, in our case the segregation is due to the combination of pressure's inhibition of movement and stochasticity. It is interesting to observe that, despite all the differences in the model, our results are in agreement with the observation of \cite{li15,morozov06,petrovskii05,petrovskii02} that this ``patchy invasion'' takes place for parameter values very close to the ones that would result in the extinction of both populations. This could be important from a therapeutic perspective, suggesting to exercise extra care when tumours' growth is subject to pressure-related effects.

In all this case, the comparison between the discrete and the continuous approach allows us to better understand which phenomena are mainly driven by stochasticity and which other can be described equally well by deterministic rules.

An important addition to the model in the future could be the explicit dynamics of oncolytic viruses. If viruses were allowed to move with standard diffusion, then it would be reasonable to expect broader infections in the model with pressure-driven cell movement, as viruses would face no restriction in moving against pressure gradient. However, it is likely that a higher cell pressure has some inhibition on viral propagation, and patterns and trend might not be too dissimilar. Similarly, it would be interesting to include in the agent-based model other biologically relevant elements that may inhibit viral delivery, such as the extracellular matrix, the influence of hypoxia or the effect of unevenly dense regions of collagen, for instance. Furthermore, the effect of an immune response when oncolytic viruses are released are still not entirely clear~\cite{Hemminki2020review}; it is important to remark that in recent years the combination of oncolytic viruses with immunotherapy has shown promising results (see \cite{engeland22} for a review of the topic).

From the mathematical point of view, a rigorous way to characterise the traveling wave solutions of the system of PDEs \eqref{eq:p} is lacking. While waves connecting $(K,0)$ to $(u^*,i^*)$ cannot be obtained starting from initial conditions in which the support of $i$ is surrounded by an area where $u=K$ (and therefore it is not possible to describe a homogeneous population invaded by a new infection or predator), our numerical simulations show that it makes sense to look for waves connecting $(0,0)$ to $(u^*,i^*)$, corresponding to the race between two expanding populations. Regarding the applications perspectives, the inclusion of viral deterministic dynamics in the stochastic model would not represent an obstacle to the derivation of the macroscopic continuum model, as the techniques used in \cite{almeida22,bubba20,macfarlane20} could be easily adapted to the resulting hybrid discrete-continuum microscopic model. However, it might be quite challenging to rigorously study a traveling wave for three populations even without the addition of other biological complications.

Finally, let us observe that all these approaches could be then useful to determine the optimal treatment protocol in different situations, both in terms of treatment schedules (as in \cite{jenner18insights,sherlock23}) and viral injection locations (as in \cite{jenner20voronoi}). Overall, there are still questions to be addressed to optimise viral delivery in oncolytic virotherapy and the balance between failure and success, as the results in our work show, is brittle. Despite some interesting achievements and some clinical progress, for example with the celebrated cases of Adenovirus H101 for neck and head cancers or T-Vec for melanomas, the goal of using viruses routinely in therapeutic setting is still elusive.

\newpage
\appendix

\section{Formal derivation of continuum models}
\label{app:derivation}

In this Appendix we describe how to derive the models discussed in the main text.

\subsection{Uninfected cells}
Uninfected cells can first move, then reproduce or die based on the pressure value and finally become infected, as explained in Section \ref{sec:model}. The principle of mass balance gives the equation
\[
\begin{split}
	u_j^{n+1}&=\Bigl[F_{j-1\to j}^n\, u_{j-1}^n+F_{j+1\to j}^n\, u_{j+1}^n+(1-F_{j\to j-1}^n-F_{j\to j+1}^n) u_{j}^n\Bigr] \\
	&\phantom{=}\times \Bigl[1+\tau G(\rho_j^n)_+-\tau G(\rho_j^n)_-\Bigl] \Bigl(1-\tau \frac{\beta}{K} i_j^n\Bigr)
\end{split}
\]
and using the algebraic relation $x_+-x_-=x$, this simplifies to
\[
\begin{split}
	u_j^{n+1}&=\Bigl[F_{j-1\to j}^n\, u_{j-1}^n+F_{j+1\to j}^n\, u_{j+1}^n+(1-F_{j\to j-1}^n-F_{j\to j+1}^n) u_{j}^n\Bigr]\\
	&\phantom{=}\times\Bigl[1+\tau G(\rho_j^n)\Bigl]\Bigl(1-\tau \frac{\beta}{K} i_j^n\Bigr)
\end{split}
\]
Let us define
\begin{equation}
	\label{eq:Xi}
	\Phi\coloneqq -(F_{j\to j-1}^n+F_{j\to j+1}^n)u_j^n+F_{j-1\to j}^n u_{j-1}^n+F_{j+1\to j}^n u_{j+1}^n
\end{equation}
so that the previous equation becomes
\begin{align*}
	u_j^{n+1}&=(u_j^n+\Phi)\Bigl[1+\tau G(\rho_j^n)\Bigl]\Bigl(1-\tau \frac{\beta}{K} i_j^n\Bigr)\\
	&=u_j^n+\tau G(\rho_j^n) u_j^n-\tau \frac{\beta}{K} u_j^n i_j^n +\Phi -\tau^2 G(\rho_j^n) \frac{\beta}{K} u_j^n i_j^n \\
	&\phantom{=}+ \tau \Phi \Bigl[ G(\rho_{j}^n)-\frac{\beta}{K} i_j^n - \tau G(\rho_{j}^n)\frac{\beta}{K} i_j^n\Bigr] 
\end{align*}
We now divide both sides of the previous equation by $\tau$ and rearrange the terms to get
\begin{equation}
	\label{eq:u_almostdone}
	\frac{u_j^{n+1}-u_{j}^n}{\tau}=G(\rho_j^n)u_{j}^n- \frac{\beta}{K} u_j^n i_j^n+\frac{1}{\tau} \Phi +H_1
\end{equation}
where
\[
H_1\coloneqq  -\tau G(\rho_j^n) \frac{\beta}{K} u_j^n i_j^n  + \Phi \Bigl[ G(\rho_{j}^n)-\frac{\beta}{K} i_j^n - \tau G(\rho_{j}^n)\frac{\beta}{K} i_j^n\Bigr] 
\]
We will shortly see that in the cases of interest $H_1$ is the sum of higher order terms and therefore vanishes as $\tau,\delta\to 0$. 

Let us now assume that there are two functions $u\in C^2([0,+\infty),\R)$ such that $u_{j}^{n}=u(t_n,x_j)=u$ and $i\in C^2([0,+\infty),\R)$ such that $i_{j}^{n}=i(t_n,x_j)=i$ (from now on we omit the arguments of functions computed at $(t_n,x_j)$); thus, we can use Taylor expansions for $u$ in time and space as follows
\begin{gather*}
	u_{j}^{n+1}=u(t_n+\tau,x_j)=u+\tau \pt u+ \bigO(\tau^2)\\
	u_{j\pm1}^{n}=u(t_n,x_j\pm\delta)=u\pm \delta \px u+ \frac{1}{2} \delta^2 \pxx u+\bigO(\delta^3)
\end{gather*}
Furthermore, we are assuming that the function $\Pi$ is smooth, so that $\rho=\Pi(u+i)\in C^2([0,+\infty),\R)$ and we can use a Taylor expansion for $\rho$ as well
\[
\rho_{j\pm1}^{n}=\rho(t_n,x_j\pm\delta)=\rho\pm \delta \px \rho+ \frac{1}{2} \delta^2 \pxx \rho+\bigO(\delta^3)
\]

Let us now treat separately the cases in which movement does or does not depend on pressure.

\paragraph{Undirected cell movement}
In this case, we have $F_{j\to j\pm 1}^n=F_{j\pm 1\to j}^n=\theta_u$, which is a constant independent of $n, j$. We then obtain
\[
\Phi=\theta_u (u_{j-1}^n+u_{j+1}^n-2u_j^n)=\theta_u \delta^2 \pxx u+\bigO(\delta^3)
\]
and thus
\[
H_1=-\tau G(\rho_j^n)\frac{\beta}{K} u_j^ni_j^n + \Bigl[ G(\rho_{j}^n)-\frac{\beta}{K} i_j^n - \tau G(\rho_{j}^n)\frac{\beta}{K} i_j^n\Bigr] \Phi= \bigO(\tau)+\bigO(\delta^2)
\]
Eq. \eqref{eq:u_almostdone} then becomes
\[
\pt u+\bigO(\tau^2)=\theta_u\frac{\delta^2}{2\tau} \pxx u +p u-\frac{\beta}{K} ui+\bigO(\tau)+\bigO(\delta^2)
\]
Letting $\tau,\delta\to 0$ in such a way that $\frac{\delta^2}{2\tau}\to D$, we obtain
\[
\pt u=\theta_u D\pxx u+pu-\frac{\beta}{K} ui
\]

\paragraph{Pressure-driven cell movement}
In this case, we modify the terms as follows:
\[
F_{j\to j\pm 1}^n\coloneqq\theta_u \frac{(\rho_j^n-\rho_{j\pm 1}^n)_+}{2P}=\frac{\theta_1 }{2P}\Bigl(\pm\delta\px \rho+\frac{1}{2}\delta^2\pxx \rho+\bigO(\delta^3)\Bigr)_+=\bigO(\delta)
\]
It is then easy to see that $H_1\to 0$ as $\tau,\delta\to 0$, as each term is multiplied either by $\tau$ or by some $F$. We then use the Taylor expansion of $u$ in Eq. \eqref{eq:Xi} to get
\[
\begin{split}
	\Phi	&=-(F_{j\to j-1}^n+F_{j\to j+1}^n)u+F_{j-1\to j}^n\Bigl(u-\delta\px u+\frac{1}{2}\delta^2\pxx u+\bigO(\delta^3)\Bigr)\\
	&\phantom{=}+F_{j+1\to j}^n \Bigl(u+\delta\px u+\frac{1}{2}\delta^2\pxx u+\bigO(\delta^3)\Bigr)\\
	&=(F_{j-1\to j}^n-F_{j\to j-1}^n+F_{j+1\to j}^n-F_{j\to j+1}^n)u+\delta(-F_{j-1\to j}^n+F_{j+1\to j}^n)\px u\\
	&\phantom{=}+\frac{1}{2}\delta^2(F_{j-1\to j}^n+F_{j+1\to j}^n)\pxx u +\bigO(\delta^3)
\end{split}
\]
Now, let us observe that
\begin{align*}
	F_{j\pm1\to j}^n-F_{j\to j\pm1}^n&=\frac{\theta_1 }{2P}[(\rho_{j\pm 1}^n-\rho_{j}^n)_+-(\rho_j^n-\rho_{j\pm 1}^n)_+]\\
	&=\frac{\theta_1 }{2P}(\rho_{j\pm 1}^n-\rho_{j}^n)=
	\frac{\theta_1 }{2P}\Bigl(\pm\delta\px \rho+\frac{1}{2}\delta^2\pxx \rho+\bigO(\delta^3)\Bigr)
\end{align*}
using the relation $x_+-(-x)_+=x_+-x_-=x$. We therefore have
\[
\Phi=\frac{\theta_u}{2P}\Bigl\{\delta^2\pxx \rho u+\delta[-(-\delta\px\rho+\bigO(\delta^2))_+ +(\delta\px\rho+\bigO(\delta^2))_+]\px u+\bigO(\delta^3)\Bigr\}
\]
Finally, Eq. \eqref{eq:u_almostdone} becomes
\begin{align*}
	\pt u+ \bigO(\tau^2)&=G(\rho)u- \frac{\beta}{K} u i+\frac{\theta_u}{P}\ \frac{\delta^2}{2\tau}\Bigl\{\pxx \rho u\\
	&\phantom{=}+[(\px\rho+\bigO(\delta))_+-(-\px\rho+\bigO(\delta))_+]\px u+\bigO(\delta)\Bigr\}+H_1
\end{align*}
Letting $\tau,\delta\to 0$ in such a way that $\frac{\delta^2}{2\tau}\to D$ we arrive at the final result:
\[
\begin{split}
	\pt u&=\frac{\theta_u D}{P}\{\pxx \rho u+[(\px\rho)_+-(-\px\rho)_+]\px u\}+G(\rho)u-\frac{\beta}{K} ui\\
	&=\frac{\theta_u D}{P}(\pxx \rho u+\px\rho\px u)+G(\rho)u-\frac{\beta}{K} ui\\
	&=\frac{\theta_u D}{P}\px (u\px\rho)+G(\rho)u-\frac{\beta}{K} ui
\end{split}
\]

\subsection{Infected cells}
Infected cells can first move, then die based on the pressure value, as explained in Section \ref{sec:model}. Also, uninfected cells may be infected. The computations follow the same strategy of uninfected cells, so we only sketch the main points. The principle of mass balance gives the equation
\begin{align*}
	i_j^{n+1}&=\Bigl[\tilde{F}_{j-1\to j}^n\, i_{j-1}^n+\tilde{F}_{j+1\to j}^n\, i_{j+1}^n+(1-\tilde{F}_{j\to j-1}^n-\tilde{F}_{j\to j+1}^n) i_{j}^n\Bigr](1-\tau q) \\
	&\phantom{=}+\tau \frac{\beta}{K} i_j^n (1+\tau G(\rho_j^n)) \Bigl[{F}_{j-1\to j}^n\, u_{j-1}^n+{F}_{j+1\to j}^n\, u_{j+1}^n+(1-{F}_{j\to j-1}^n-{F}_{j\to j+1}^n) u_{j}^n\Bigr]
\end{align*}
which simplifies to
\begin{align*}
	i_j^{n+1}&=(1-\tau q)\Bigl[\tilde{F}_{j-1\to j}^n\, i_{j-1}^n+\tilde{F}_{j+1\to j}^n\, i_{j+1}^n+(1-\tilde{F}_{j\to j-1}^n-\tilde{F}_{j\to j+1}^n)i_{j}^n\Bigr]\\
	&\phantom{=} +\tau \frac{\beta}{K} u_j^n i_j^n+\tau H_2\\
	&=\Bigl[\tilde{F}_{j-1\to j}^n\, i_{j-1}^n+\tilde{F}_{j+1\to j}^n\, i_{j+1}^n-(\tilde{F}_{j\to j-1}^n+\tilde{F}_{j\to j+1}^n)i_{j}^n\Bigr]+(1-\tau q)i_j^n\\
	&\phantom{=}+\tau \frac{\beta}{K} u_j^n i_j^n+\tau H_2+\tau H_3 \\
	&=\Psi +(1-\tau q)i_j^n+\tau \frac{\beta}{K} u_j^n i_j^n+\tau H_2+\tau H_3
\end{align*}
where
\[
\Psi\coloneqq \tilde{F}_{j-1\to j}^n\, i_{j-1}^n+\tilde{F}_{j+1\to j}^n\, i_{j+1}^n-(\tilde{F}_{j\to j-1}^n+\tilde{F}_{j\to j+1}^n)i_{j}^n
\]
and
\begin{gather*}
	\begin{split}
		H_2&\coloneqq \tau  G(\rho_j^n) \frac{\beta}{K} u_j^n i_j^n +\frac{\beta}{K} i_j^n (1+\tau G(\rho_{j}^n))\\
		&\phantom{=}\times\underbrace{\Bigl[{F}_{j-1\to j}^n\, u_{j-1}^n+{F}_{j+1\to j}^n\, u_{j+1}^n-({F}_{j\to j-1}^n+{F}_{j\to j+1}^n) u_{j}^n\Bigr]}_{=\Phi}
	\end{split}
	\\
	H_3\coloneqq -q\underbrace{\Bigl[\tilde{F}_{j-1\to j}^n\, i_{j-1}^n+\tilde{F}_{j+1\to j}^n\, i_{j+1}^n-(\tilde{F}_{j\to j-1}^n+\tilde{F}_{j\to j+1}^n)i_{j}^n\Bigr]}_{=\Psi}
\end{gather*}
Dividing both sides by $\tau$ and rearranging the terms we get 
\begin{equation}
	\label{eq:i_almostdone}
	\frac{i_j^{n+1}-i_j^n}{\tau}=\frac{1}{\tau}\Psi-qi_j^n+\frac{\beta}{K} u_j^n i_j^n+H_2+H_3
\end{equation}

Let us now make the same regularity assumptions of the previous subsection and use the Taylor expansions for $i$ to arrive at
\begin{gather*}
	i_{j}^{n+1}=i(t_n+\tau,x_j)=i+\tau \pt i+ \bigO(\tau^2)\\
	i_{j\pm1}^{n}=i(t_n,x_j\pm\delta)=i\pm \delta \px i+ \frac{1}{2} \delta^2 \pxx i+\bigO(\delta^3)
\end{gather*}
Again, we treat separately the case in which movement does not depend on pressure and the case in which it does.

\paragraph{Undirected cell movement}
In this case, we have
\[
\Phi=\theta_u (u_{j-1}^n+u_{j+1}^n-2u_j^n)=\theta_u \delta^2 \pxx u+\bigO(\delta^3)
\]
and
\[
\Psi=\theta_i (i_{j-1}^n+i_{j+1}^n-2i_j^n)=\theta_i \delta^2 \pxx i+\bigO(\delta^3)
\]
This means that $H_2+H_3=\bigO(\tau)+\bigO(\delta^2)$. Eq. \eqref{eq:i_almostdone} then becomes
\[
\pt i+\bigO(\tau^2)=\theta_i\frac{\delta^2}{2\tau} \pxx i +\frac{\beta}{K} ui-qi+\bigO(\tau)+\bigO(\delta^2)
\]
Letting $\tau,\delta\to 0$ in such a way that $\frac{\delta^2}{2\tau}\to D$ we obtain the required term:
\[
\pt i=\theta_i D\pxx i+\frac{\beta}{K} ui-qi
\]

\paragraph{Pressure-driven cell movement}
In this case $F_{j\to j\pm 1}^n$ and $\tilde{F}_{j\pm 1\to j}^n$ are $\bigO(\delta)$. 
It is then easy to see that $H_2+H_3\to 0$ as $\tau,\delta\to 0$, as each term is multiplied either by $\tau$, by some $F$ or by some $\tilde{F}$.
We can then repeat the calculations already performed for the uninfected cells to show that, letting $\tau,\delta\to 0$ in such a way that $\frac{\delta^2}{2\tau}\to D$ yields
\[
\frac{1}{\tau}\Psi \to \frac{\theta_i D}{P}\px (i\px\rho)
\]
Therefore, in the required limit Eq. \eqref{eq:i_almostdone} becomes
\[
\pt i=\frac{\theta_i D}{P}\px (i\px\rho)+\frac{\beta}{K} ui-qi
\]

\section{Details of numerical simulations}
\label{app:num}

\paragraph{Parameter values}
In Table \ref{tab:parameters} we list the parameters we adopt as  a reference in the numerical simulations. Some simulations use other parameter values to explore different behaviours emerging from our model, as explained in the main text.

Most of the parameters in the model have been estimated from the existing experimental literature. The maximal duplication rate in the logistic growth $p$ has been taken equal to $\log(2)/37\;$h$^{-1}\approx 1.87 \times 10^{-2}\;$h$^{-1}$; the duplication time of 37 hours is approximately the highest among the ones reported in \cite{ke00} and has been chosen so that the exponential growth is not too fast. The death rate of infected cells $q$ has been taken equal to $1/24\;$h$^{-1}=4.17\times 10^{-2}\;$h$^{-1}$, following \cite{ganly00}.

The diffusion coefficient of uninfected cells $D_u$ has been estimated from the experimental data of the U343 control group of \cite{kim06}, as already done in \cite{pooladvand21}: the tumour volume passes in 40 days from $70\;$mm$^3$ to $1000\;$mm$^3$, which corresponds to a change in the tumour radius from approximately $2.6\;$mm to approximately $6.2\;$mm; since in absence of viral infection the dynamic of uninfected cells follows Eq. \eqref{eq:fisher} in the case of undirected movement and Eq. \eqref{eq:nonlin_diff} in the case of pressure-driven movement, we can estimate the diffusion coefficient from the wave speed formulas described in Section \ref{sec:wave} and obtain
\[
D_u=\frac{c^2}{4p}=\Bigl(\frac{6.2-2.6\;\text{mm}}{40\times 24\;\text{h}}\Bigr)^2 \times \frac{1}{4 \times 1.87 \times 10^{-2}\,\text{h}} \approx 1.88 \times 10^{-4}\;\text{mm}^2/\text{h}
\]
in the former case and 
\[
D_u=\frac{2c^2}{p}\approx 8 \times 1.88 \times 10^{-4}\;\text{mm}^2/\text{h}\approx 1.50 \times 10^{-3}\;\text{mm}^2/\text{h}
\]
in the latter case. We assume $D_i=D_u$, as a priori we have no reason to believe that the infection affects cellular movement.

In the agent based models, cellular movement is governed by the parameters $\theta_u$ and $\theta_i$. Having in mind the formal derivation of the continuum models, we set them according to the formula
\begin{equation}
	\label{eq:theta}
	\theta_k=
	\begin{cases}
		\dfrac{2\tau D_k}{\delta^2} \quad&\text{in one dimension}\\[8pt]
		\dfrac{4\tau D_k}{\delta^2} \quad&\text{in two dimensions}
	\end{cases}
\end{equation}
where $k=u,i$ and $\tau, \delta$ are the temporal and spatial discretisations. For the sake of simplicity, in the text we always refer to the the variation of the diffusion coefficients (which have a clear macroscopic meaning), keeping in mind that $\theta_u$ and $\theta_i$ are adjusted accordingly.

The carrying capacity $K$ has been estimated assuming that a cell has radius $10\;\mu$m$=10^{-2}\;$mm \cite[\S1.1]{lodish08}: this implies that  the carrying capacity is $10^2\;$cells/mm in one spatial dimension and $10^4\;$cells/mm$^2$ in two spatial dimensions. Since we observed that $K=100\;$cells/mm is too little to obtain a good agreement between agent-based and continuum models, we decided to increase it to $K=1000\;$cells/mm in the case of one spatial dimension.

Let us now estimate the infection rate $\beta$. In \cite{friedman06} the authors assume that oncolytic viruses have an infection rate of $\hat{\beta}=7\times 10^{-10}\;$mm$^3$/(viruses$\times$h) and cells have a carrying capacity of $\hat{K}=10^6\;$cells/mm$^3$. Clearly, this value cannot directly be used in our model, which does not take into account viral dynamics explicitly. Let us assume that viral density satisfies the PDE
\[
\pt v(t,x)=D_v\pxx v(t,x)+\alpha q i(t,x)-q_v v(t,x)
\]
where $D_v$ is the diffusion coefficient, $\alpha$ is the number of viruses released by the lysis when an infected cell dies, $q$ is the death rate of infected cells (which is also present in our model) and $q_v$ is the clearance rate of the virus. Since viral dynamics are faster than cellular dynamics, we can assume that the viral density is quasi-steady, leading to the algebraic relation
\begin{equation*}
	v(t,x)=\frac{\alpha q}{q_v}\ i(t,x)
\end{equation*}
We therefore have $\beta=\hat{\beta} \alpha q \hat{K}/q_v$, with the values of the parameters $\alpha$ and $q_v$ to be chosen. We set $q_v=1/6\;$h$^{-1}$ as in \cite{mok09}. The viral load released by the death of infected cells depends highly on the type of virus and ranges from the value $157\pm 23.4\;$viruses/cell estimated in \cite{workenhe14} to the value $3500\;$viruses/cell of \cite{chen01}; we chose an intermediate value of $\alpha=580\;$viruses/cell. In conclusion, we have
\[
\beta= \Bigl(7\times 10^{-10}\;\frac{\text{mm}^3}{\text{viruses}\times\text{h}}\Bigr) \times \frac{580\;\text{viruses/cell}\times 1/24\;\text{h}^{-1}}{1/6\;\text{h}^{-1}} \times 10^6\; \frac{\text{cells}}{\text{mm}^3} \approx 0.102\; \text{h}^{-1}
\]
It is important to observe that the parameter $\beta$ incorporates a wide variety of dynamics related to the virus, hence different parameter values could result in the same value of $\beta$: for example, if we assume a faster viral decay of $q_v=1\;$h$^{-1}$ and a higher number of viruses released during lysis $\alpha=3500\;$viruses/cell as in \cite{chen01}, then the value of $\beta$ remains unchanged.

Finally, we use the initial conditions given in Eq. \eqref{eq:initial}. The value of $R_u$ has been set to $2.6\;$mm, as in \cite{kim06}. Remembering that we postulate a central injection \cite{russell18}, we observe that it is not easy to find reliable estimates for the radius of the region occupied by infected cells right after such an injection. We thus chose the value $R_i=1\;$mm, which is slightly less than half of the tumour radius. In the model with undirected movement, initial conditions do not really affect long-time dynamics of the system. On the other hand, varying initial conditions in the model with pressure-driven movement may result in opposite outcomes, as the infection is unable to propagate in areas of constant total density. This is yet another indication of the sensitivity of the process when a tumour's growth is strongly influenced by the pressure.

Numerical simulations are run until the final time $T=1500\;$h, since their behavior up to this moment is also representative of later dynamics. For the spatial domain $[-L,L]$ (or $[-L,L]^2$) we set $L=10\;$mm so that wave fronts do not hit the boundary before $T$ and the domain is representative of typical extensions of solid tumours. Since some simulations of the model with pressure-driven movement have been performed with a higher value of $D_u$, $D_i$ (and, consequently, a higher speed wave) in this situations $L$ has been slightly increased.

\paragraph{Numerical simulations for the discrete models}

We used a temporal step $\tau=0.02\;$h and a spatial step of $\delta=0.1\;$mm both for the one-dimensional and the two-dimensional simulations. Some additional simulations (not shown) demonstrated that a further refinement on the grid does not result in a significant improvement of the agreement between discrete and continuous models, while greatly affecting the computation time. All simulations have been performed in \textsc{Matlab 2021b}.

At every iteration we first computed the sum of the two populations and then cell numbers are updated according to the rules described in Section \ref{sec:model}. We first consider movement, then reproduction and death, finally infection. Zero-flux boundary conditions are implemented by not allowing cells at the boundary to leave the domain. This however does not have any particular influence on the results, as all the simulations are stopped before the wave front reaches the boundary.

Observe that the sum of the two populations is not updated during the iteration, in accordance with the formulas of Section \ref{sec:model}. As a consequence, cell densities may fluctuate above the carrying capacity for short periods of time, before the dynamics makes them decrease again. In order to avoid problems with the formula in Eq. \eqref{eq:movement_pressure}, we truncate the pressure at the carrying capacity $K$. In the case of undirected movement, the fluctuations above the carrying capacity are more frequent because of the lack of movement inhibition in crowded regions (see Fig. \ref{fig:l_oned}); this however does not cause any problem in the formulation of the model. 

Since we only need to keep track of the collective fate of cells in the same lattice point, we used the built-in \textsc{Matlab} functions \texttt{binornd} and \texttt{mnrnd}, which compute random arrays according to binomial and multinomial distributions. 

Figs. \ref{fig:l_oned}, \ref{fig:l_twod}, \ref{fig:n_varie}a-b, \ref{fig:p_oned_fail}, \ref{fig:p_twod_fail} and \ref{fig:p_twod_sucq} show the average of five simulations. On the other hand, Figs. \ref{fig:l_varie}, \ref{fig:n_varie} (with the exception of Figs. \ref{fig:n_varie}a-b) and \ref{fig:p_varie} have the purpose of explaining the influence of stochastic effects and therefore show a single simulation. Averaging simulations in two dimensions results in nonzero cell densities below $\frac{1}{\delta^2}$, which makes no sense in the case of a single simulation; for the sake of consistence, we decided to truncate these values to zero.

In order to allow reproducibility, a random seed has been set at the beginning of each new simulation, ranging from 1 to 5. In the figures representing a single simulation only the one with random seed equal to 1 is shown.

\paragraph{Numerical simulations for the continuum models}
Eqs. \eqref{eq:l} and \eqref{eq:p} complemented with homogeneous Neumann boundary conditions have been solved in \textsc{Matlab 2021b} using the built-in function \texttt{pdepe}; in the two-dimensional case we exploited the radial symmetry of the equations. We considered a uniform discretisation of the spatial interval $[0,L]$ consisting of $500$ points and a uniform discretisation of step $1$ of the temporal interval $[0,T]$.

The application of this strategy to the simulation depicted in Figs. \ref{fig:p_oned_fail}a, \ref{fig:p_twod_fail}a caused some numerical instabilities. We solved this issue by using a forward upwind scheme for the transport term, following \cite{leveque07}. This method is able to deal with such instabilities even at long times if we take the discretisation $\Delta x=0.05$, $\Delta t=10^{-4}$. We then used the same algorithm for all the numerical solutions of Eq. \eqref{eq:p} for the sake of coherence, with discretisations $\Delta x=0.1$, $\Delta t=10^{-4}$. In the two-dimensional case, we can rely upon the radial symmetry of the problem; hence, the analog of Eq. \eqref{eq:p} becomes

\begin{equation*}
	\begin{cases}
		\pt u=\dfrac{D_u}{K} \dfrac{1}{r} \pr [r\,u\pr (u+i)]+pu\Bigl(1-\dfrac{u+i}{K}\Bigr)-\dfrac{\beta}{K} ui \\[8pt]
		\pt i=\dfrac{D_i}{K} \dfrac{1}{r} \pr [r\,i\pr (u+i)]+\dfrac{\beta}{K} ui-qi
	\end{cases}
\end{equation*}

In the two-dimensional plots of the supplementary material we truncated the solutions at a value $\frac{1}{\delta^2}$ to be consistent with the representation of the agent-based model. We also use the same threshold $\frac{1}{\delta^2}$ to identify the wave front, which is depicted in some of the Figures. However, it is important to observe that in some simulations the density of infected cells is positive almost in the whole domain occupied by uninfected cells; the definition of the front as the location in which cell density is above $\frac{1}{\delta^2}$ makes visual comparison between the discrete and the continuous model clearer.

\bibliographystyle{siam}
\bibliography{bibliography_rev}
	
\end{document}